\RequirePackage{ifpdf}
\documentclass[11pt,a4paper]{article}
\pdfoutput=1 

\usepackage{amsmath} 
\usepackage{jheppub}
\usepackage{pstricks}
\usepackage{longtable}
\usepackage[final]{pdfpages}
\usepackage{ifpdf}
\usepackage{flexisym}
\usepackage{breqn}
\usepackage{slashed}

\def\M{{\cal M}}
\def\ep{\epsilon}
\def\unM{\hat{\cal M}}
\def\unas{ \left( \frac{\hat{a}_s}{\mu_0^{\epsilon}} S_{\epsilon} \right) }
\def\rnM{{\cal M}}
\def\rnas{ \left( a_s  \right) }
\def\b0{\beta_0}
\def\cD{{\cal D}}

\def\spt{(s+t)}
\def\spu{(s+u)}
\def\tpu{(t+u)}

\title{Two-Loop QCD Corrections to Higgs $\rightarrow b + \bar{b} + g$ Amplitude}

\author[a]{Taushif Ahmed,}
\author[a]{Maguni Mahakhud,}
\author[b]{Prakash Mathews,}
\author[a]{Narayan Rana}
\author[c]{and V. Ravindran}

\affiliation[a]{Regional Centre for Accelerator-based Particle Physics,\\ 
Harish-Chandra Research Institute, Chhatnag Road, Jhunsi, Allahabad 211 019, India}
\affiliation[b]{Saha Institute of Nuclear Physics, 1/AF Bidhan Nagar, Kolkata 700 064, India}
\affiliation[c]{The Institute of Mathematical Sciences, C.I.T Campus, 4th Cross St, Tharamani, Chennai 600 113, India}

\emailAdd{taushif@hri.res.in}
\emailAdd{maguni@hri.res.in}
\emailAdd{prakash.mathews@saha.ac.in}
\emailAdd{narayan@hri.res.in}
\emailAdd{ravindra@imsc.res.in}

\abstract{
Exclusive observables involving Higgs boson in association with jets 
are often well suited to study the Higgs boson properties. 
They are rates involving cuts on the final state jets or differential
distributions of rapidity, transverse momentum of the observed Higgs boson.
While they get dominant contributions from gluon initiated partonic
subprocesses,  it is important to include the subdominant ones coming from other
channels.   In this article, we study one such channel namely the Higgs production in association
with a jet in bottom anti-bottom annihilation process. 
We compute relevant amplitude $H\rightarrow b+\overline b+g$ up to two loop level in QCD
where Higgs couples to bottom quark through Yukawa coupling.
We use projection operators to obtain
the coefficients for each tensorial structure appearing in this process.
We have demonstrated that the renormalized amplitudes do have the right infrared
structure predicted by the QCD factorization in dimensional regularization.  
The finite parts of the one and two loop amplitudes
are presented after subtracting the infrared poles using Catani's subtraction operators.  
} 
\preprint{HRI-RECAPP-2014-010}

\begin{document}
\unitlength1cm
\maketitle
\flushbottom

\section{Introduction}
\setcounter{equation}{0}
\label{sec:intro}
The tests of the Standard Model (SM) have been going on for several decades
in various experiments and most of its predictions have been tested in an 
unprecedented accuracy.
The recent discovery of Higgs boson by ATLAS \cite{atlashiggs} and CMS \cite{cmshiggs} collaborations 
at the Large Hadron Collider (LHC) puts 
the SM on firm footing.  The Higgs boson results from Higgs mechanism
that provides a framework for electroweak symmetry breaking.  Elementary particles
such as leptons, quarks, gauge bosons and Higgs boson acquire 
masses through the Higgs mechanism.  The mass of the Higgs boson being a parameter of the
theory can not be predicted by the SM and hence 
its discovery provides a valuable information on this.   
Results from Higgs searches at LEP \cite{higgslep} and Tevatron \cite{higgstev} were crucial ingredients
to the recent discovery in narrowing down the search regions for the LHC
collaborations.  The direct searches at the LEP excluded Higgs of mass below 114.4 GeV 
and the precision electroweak measurements \cite{lepprecis} 
hinted for Higgs boson in the mass less than 152 GeV at $95\%$ confidence level (CL).
Tevatron on the other hand excluded Higgs of mass in the range $162-166$ GeV at $95\%$ CL.   

The dominant production mechanism for the Higgs production at the LHC 
is gluon gluon fusion through top quark loop.  
The subdominant ones 
come from vector boson fusion, associated production of
Higgs with vector bosons and top anti-top pairs and bottom anti-bottom annihilation.
The inclusive production cross section for the Higgs production 
is known to an unprecedented accuracy 
due to many breakthroughs in the computation of amplitudes, loop and phase space integrals.
For gluon-gluon \cite{gghNNLO}, vector boson fusion processes \cite{bolzoni}, and associated production with vector
bosons \cite{Han:1991ia}, the inclusive rates are known to NNLO accuracy in QCD. 
There are also studies related to the 
Higgs production in association with bottom quarks which were also motivated to
study Higgs boson in certain SUSY models, namely MSSM.
The coupling of bottom quarks become large in the large $\tan\beta$ region, where
$\tan\beta$ is the vacuum expectation values of up and down type Higgs fields in
the Higgs sector of MSSM.  Such large couplings can enhance gluon fusion as well as
bottom quark fusion subprocesses.  
Fully inclusive cross section for Higgs production in association with bottom quark  
to NNLO level accuracy is also known in the variable flavour scheme (VFS) \cite{vfs}, while it is known
only up to NLO level in the fixed flavour scheme (FFS) \cite{ffs}.   In the VFS, one assumes the initial
state bottom quarks inside the proton.  They are there as a result of emission of 
collinear bottom anti-bottom states from the gluons intrinsically present inside the proton.  
They being collinear give
large logs which need to be resummed.  The resummed
contribution is the source for non-vanishing bottom and anti-bottom parton distribution
functions inside the proton in the VFS scheme.

The differential distributions for Higgs production and its decay to pair of photons \cite{Anastasiou:2005qj} or
massive vector bosons \cite{Anastasiou:2007mz, Grazzini:2008tf} have also been known at NNLO level in QCD in 
the infinite top quark mass limit.
Such exclusive observables allow direct comparison of theoretical predictions with 
experimental results which include kinematical cuts on the final state particles.
In particular, observables with jet vetos enhance the significance of the
signal considerably allowing us to study the properties of Higgs boson and its
coupling to other SM particles.  NNLO QCD prediction \cite{Boughezal:2013uia} for production of Higgs with one jet
through effective gluon-gluon-higgs vertex in the infinite top quark mass limit is available,
thanks to various ingredients that are computed to the 
required accuracy by different groups \cite{Hggg, Gehrmann:2011aa}.
As the experimental accuracy improves, it will be important to include other subdominant 
production mechanisms.  In this article, we provide the relevant one and two loop amplitudes
for the process $H \rightarrow b + \overline b + g $ which is analytically continued
also to obtain the production of Higgs boson with one jet in 
bottom anti-bottom annihilation, i.e., $b+\overline b \rightarrow H+g$, where Higgs couples
to bottom quark through Yukawa coupling denoted by $\lambda$.  We use VFS scheme throughout.
This will be an important supplement to the Higgs boson with one jet at NNLO level as it includes
the bottom quark effects in VFS scheme. 

Beyond leading order in perturbation theory, one encounters large number of
Feynman amplitudes with rich Lorentz and gauge structures.
In addition, the loop integrals become increasingly complicated due to their 
multiple kinematic dependence.  
Generation of diagrams, simplification of Lorentz, Dirac and
color indices can be done symbolically.  
Using integration by parts (IBP) and Lorentz invariant (LI) identities
the large number of loop integrals can be reduced in a rather
straight forward way to few master integrals (MI).  
The two loop MIs for four legs processes where all fields but one external
leg are massless were solved by Gehrmann and Remiddi \cite{Gehrmann:2000zt} 
using an elegant method of differential equations.  

In this article we present one and two loop QCD amplitudes for the process 
$H \rightarrow b + \overline b + g $ treating both bottom and other four light quarks massless.  
We do not include top quark in our analysis.  
To obtain infrared safe observables, 
we require, in addition to these two loop amplitudes, 
one loop corrected 
$H\rightarrow b+\overline b + 2~{\rm partons }$ and 
tree level $H \rightarrow b+\overline b + 3~{\rm partons}$ amplitudes.
Note that they are individually infrared singular due to the presence of massless partons
in the amplitudes.   
There exist several equally
efficient frameworks which use these infrared sensitive contributions
to combine them to obtain infrared safe observables.  They go by the names 
sector decomposition \cite{secdec}, $q_T$-subtraction \cite{qtsub} and antenna subtraction \cite{antsub} methods.
More recently the method developed by Czakon using sector decomposition and FKS \cite{fks} phase
space slicing, was applied to obtain top quark pair production \cite{topnnlo} at NNLO level and
NNLO QED corrections \cite{nnloqed} to $Z\rightarrow e^+ e^-$.  Antenna subtraction was used to
obtain NNLO QCD corrections to di-jet production at the LHC.  
The NNLO corrections to Higgs plus one jet resulting
from only gluon-gluon-Higgs effective interaction are obtained recently in \cite{Gehrmann:2011aa}
making best use of the subtraction methods in an efficient way.
The amplitudes presented in this article will constitute contributions
coming from bottom-antibottom-higgs interactions to Higgs plus one jet observable
at NNLO level.  We have presented the amplitudes 
in the form suitable for easier implementation to study infra-red safe 
hadron level observables involving Higgs plus one jet at NNLO in QCD. 

In the next section, we discuss the Lagrangian that describes coupling of
Higgs boson with bottom quark, explain how the projector technique can be used to obtain the amplitudes and
describe the renormalization and factorization properties of the amplitudes.  Section \ref{sec:calc} is dedicated to the computational
details.  Final results in compact form are given in Section \ref{sec:result} and corresponding coefficients are given in the Appendix.  
In section \ref{sec:conc}, we conclude with
our findings. 

\section{Theory}
The interaction part of the action involving bottom quarks and Higgs
boson is given by 
\begin{eqnarray}
S^b_{I} = - \lambda \int d^4 x \, \phi(x) \overline \psi_b(x) \psi_b(x)
\end{eqnarray}
where, $\psi_b(x)$ denotes the bottom quark field and $\phi(x)$ the scalar field.
$\lambda$ is the Yukawa coupling given by $\sqrt{2} m_b/\upsilon$, with 
the bottom quark mass $m_b$ and 
the vacuum expectation value $\upsilon \approx 246$ GeV.  For the pseudoscalar Higgs 
of MSSM, we need to replace $\lambda \phi(x) \overline \psi_b(x) \psi_b(x)$ by 
$\tilde \lambda \tilde \phi(x) \overline \psi_b(x) \gamma_5 \psi_b(x)$ in the above
equation.  The MSSM couplings are 
\[
 \tilde{\lambda} = \left\{
  \begin{array}{ll}
    -  \frac{\sqrt{2} m_b \sin\alpha}{\upsilon \cos\beta}  \,,& \qquad \tilde{\phi} = h\,,\\
    \phantom{-}  \frac{\sqrt{2} m_b \cos\alpha}{\upsilon \cos\beta}  \,,& \qquad \tilde{\phi}=H\,,\\
    \phantom{-}  \frac{\sqrt{2} m_b \tan\beta}{\upsilon } \,, & \qquad \tilde{\phi}=A\,
  \end{array}
  \right.
\]
respectively.  The angle $\alpha$ is the measure of mixing of weak and mass eigenstates
of neutral Higgs bosons. In the VFS scheme, except in the Yukawa coupling,
$m_b$ is taken to be zero like other light quarks in the theory.  The number of
active flavours is taken to be $n_f=5$.  We work in Feynman gauge throughout.  

\subsection{Notation and kinematics}
\label{subsec:defs}
\setcounter{equation}{0}
We consider the decay of Higgs boson to a bottom quark, anti-bottom quark and a gluon
\begin{equation}
H(q) \longrightarrow b(p_1) + \bar{b} (p_2) + g(p_3) \, .
\end{equation}
The associated Mandelstam variables are defined as
\begin{equation}
 s \equiv (p_1 + p_2)^2, \hspace{1cm} t \equiv (p_2 + p_3)^2, \hspace{1cm} u \equiv (p_1 + p_3)^2
\end{equation}
which satisfy 
\begin{equation}
 s >0,~ t > 0, ~ u > 0, ~~ s + t + u = M_H^2 \equiv Q^2 > 0
\end{equation}
where, $M_H$ is the mass of the Higgs boson.  We also define the following
dimensionless invariants which appear in harmonic polylogarithms (HPL) 
\cite{remiddi} and 2dHPL \cite{Gehrmann:2000zt} as 
\begin{equation}
 x \equiv s/Q^2, \hspace{1cm} y \equiv u/Q^2, \hspace{1cm} z \equiv t/Q^2
\end{equation}
satisfying  
\begin{equation}
 0 < x < 1,~ 0 < y < 1,~ 0 < z < 1,~ ~\text{and}~ x + y + z = 1.
\end{equation}

\subsubsection*{Analytical continuation} \label{sec:ancont}
In order to compute the Higgs + 1 jet production at hadron colliders, the decay amplitudes
must be analytically continued to the appropriate kinematical regions.
The corresponding processes are 
\begin{align}
  \nonumber
  1. ~~~~ \overline{b} (-p_1) + b(-p_2) &\rightarrow g(p_3) + H (p_4) 
  \\ \nonumber
  2. ~~~~ b (-p_2) + g(-p_3) &\rightarrow b (p_1) + H (p_4)
  \\ \label{subprocess}
  3. ~~~~ \overline{b}(-p_1) + g(-p_3) &\rightarrow \overline{b} (p_2) + H (p_4)
\end{align}

For the process 1,  
% $p_1 + p_2 \rightarrow p_3 + p_4$, 
$Q^2 = M_H^2 > 0$, $s > 0 ,~ t < 0 $ and $u < 0$. Hence we introduce the dimensionless
parameters $u_1$ and $v_1 $ with the following definitions 
\begin{equation}
 u_1 \equiv - \frac{u}{s}, \hspace{1cm} v_1 \equiv \frac{Q^2}{s}
\end{equation}
such that $0 < u_1 < 1$ and $0< v_1 < 1$.

Similarly, for the process 2,
% $p_2 + p_3 \rightarrow p_1 + p_4$, 
$Q^2 = M_H^2 > 0$, $s < 0 ,~ t > 0 $ and $u < 0$ and the dimensionless
parameters are $u_2$ and $v_2 $ with the following definitions 
\begin{equation}
 u_2 \equiv - \frac{u}{t}, \hspace{1cm} v_2 \equiv \frac{Q^2}{t}
\end{equation}
such that $0 < u_2 < 1$ and $0< v_2 < 1$.
The last one is trivially related to the second one.

\subsection{The general structure of the amplitude}
\label{sec:gentensor}
In this section, we describe how the amplitude for 
$H \rightarrow b+\overline b + g$ can be obtained using projector technique. 
Since the amplitude contains one external gluon, it  
can be expressed as 
\begin{equation}
| \M \rangle = {\cal S}_{\mu} (b,\bar{b};g) \varepsilon^{\mu}
\end{equation}
where, $\varepsilon^{\mu}$ is the gluon polarization vector. 

We observe the amplitude has the following general structure
in terms of the coefficients $A',A''$ and $A_2$:
\begin{equation}
{\cal S}_{\mu} (b,\bar{b};g) = \bar{u} (p_1) \Big\{
A'~ p_{1 \mu} + A''~ p_{2 \mu} + A_2~\slashed p_3 \gamma_{\mu} 
\Big\} v (p_2)
\end{equation}
where, we have used $p_3.\varepsilon = 0$. QCD Ward identity gives 
\begin{equation}
 A'~ p_{1}. p_3 + A''~ p_{2} . p_3 = 0 
 ~~ \Rightarrow ~~
 A' = - A''~ \frac{p_{2} . p_3}{p_{1}. p_3} \equiv A_1~ p_2.p_3 \, .
\end{equation}
Hence, the amplitude takes the following form:
\begin{align} 
{\cal S}_{\mu} (b,\bar{b};g) ~\varepsilon^{\mu} &= \bar{u} (p_1) \Big\{
A_1~( p_2.p_3 ~ p_{1 \mu} - p_1 . p_3 ~ p_{2 \mu} ) 
+ A_2~\slashed p_3 \gamma_{\mu} 
\Big\} v(p_2) ~\varepsilon^{\mu}
\nonumber\\ \label{tencoeff}
&\equiv A_1~ {\rm T_1} + A_2~ {\rm T_2} \; .
\end{align}
The coefficients $A_m ~ (m = 1, 2)$ can be obtained from the amplitude $| \M \rangle$  
using appropriate projectors ${\cal P}(A_m)$
\begin{equation}
A_m=\sum_{\rm spins} {\cal P} (A_m) {\cal S}_{\mu} (b,\bar{b};g) ~\varepsilon^{\mu}  =\sum_{\rm spins} {\cal P} (A_m) | \M \rangle
\end{equation}
where, in $d$ space-time dimensions, the projectors are found to be 
\begin{align} \label{projectors}
{\cal P} (A_1) &= \frac{2 (d-2)}{s^2 \, t \, u \, (d-3)} {\rm T_1}^{\dagger}
                + \frac{1}{s \, t \, u \, (d-3)} {\rm T_2}^{\dagger} \, ,
\nonumber\\
{\cal P} (A_2) &= \frac{1}{s \, t \, u \, (d-3)} {\rm T_1}^{\dagger}  
                + \frac{1}{2 \, t \, u \, (d-3)} {\rm T_2}^{\dagger} \, .       
\end{align}
Expanding the coefficients $A_m$ in powers of strong coupling constant $a_s = g_s^2/16 \pi^2$, we
obtain 
\begin{equation} \label{coeffexpnd}
A_m = \frac{\lambda}{\mu_R^\epsilon} ~4\pi \sqrt{a_s} T^a_{ij} \Big\{
A_m^{(0)} + a_s A_m^{(1)}
+ a_s^2 A_m^{(2)} + {\cal O}(a_s^3)
\Big\}
\end{equation}
where, $T^a$ are the Gell-Mann matrices, $a$ is adjoint and $i$, $j$ are fundamental indices of SU(3) 
and $\mu_R$ is the renormalization scale.
These coefficients $A_m^{(l)}$ completely specify the amplitude order by order in perturbation theory.

As described in section \ref{sec:ancont}, for Higgs + 1 jet production, the above amplitudes have to be 
suitably crossed and the coefficients $A_{m}$ will be expressed in terms of corresponding $u_i$ and $v_i$.

\subsection{Ultraviolet renormalization}
The Feynman amplitudes for the process $H\rightarrow b+\overline b+g$ beyond leading order
develop ultraviolet divergences in QCD.
We have used dimensional regularization to regulate them taking space-time dimension to be 
$d=4+\epsilon$. 
The scale $\mu_0$ is introduced to scale the mass dimension of the dimension-full
strong coupling constant in $d$ dimensions.   If we denote the dimensionless strong
coupling constant by $\hat g_s$ in $d$ dimensions, then the unrenormalized
amplitude can be expanded in terms of $\hat a_s=\hat g_s^2/16 \pi^2$ as 
\begin{equation} \label{unm}
 |{\cal M} \rangle = \frac{\hat{\lambda}}{\mu_0^\epsilon} S_\epsilon \unas^{\frac{1}{2}} \left\{  |\unM^{(0)} \rangle + \unas |\unM^{(1)} \rangle + \unas^2 |\unM^{(2)} \rangle + {\cal O}(\hat{a}_s^3) \right\} 
\end{equation}
where, $S_{\ep} = \exp[\frac{\ep}{2} (\gamma_E - \ln 4\pi)]$ with Euler constant $\gamma_E = 0.5772 \ldots$ , results from loop integrals beyond leading order.
$|\unM^{(i)} \rangle$ is the unrenormalized color-space vector which represents the $i^{th}$ loop amplitude.
In $\overline {MS}$ scheme, the renormalized coupling constant $a_s \equiv a_s (\mu_R^2)$ at the
renormalization scale $\mu_R$ is related to unrenormalized coupling constant $\hat{a}_s$ by 
\begin{align} \label{renas}
 \frac{\hat{a}_s}{\mu_0^{\epsilon}} S_{\epsilon} &= \frac{a_s}{\mu_R^{\epsilon}} {Z}(\mu_R^2)
\nonumber\\[1ex]
 &= \frac{a_s}{\mu_R^{\epsilon}} \left[   1 + a_s \left( \frac{1}{\ep} r_{a_{1;1}} \right) 
  + a_s^2 \left( \frac{1}{\ep^2} r_{a_{2;2}} + \frac{1}{\ep} r_{a_{2;1}} \right) + {\cal O}(a_s^3)  \right] 
\end{align}
where,
\begin{equation*}
 r_{a_{1;1}} = 2 \b0 \;, \quad
 r_{a_{2;2}} = 4\b0^2 \;, \quad
 r_{a_{2;1}} = \beta_1 \;, \quad
\end{equation*}

\begin{equation}
 \b0 = \left(\frac{11}{3} C_A - \frac{4}{3} T_F n_f\right) , \hspace{0.5cm}
 \beta_1 =  \left(\frac{34}{3} C_A^2 - \frac{20}{3} C_A T_F n_f - 4 C_F T_F n_f\right) 
\end{equation}
with $C_A = N$, $C_F = (N^2 -1)/{2 N} $, $T_F = 1/2$ and $n_f$ is the number of active quark flavors.  
The bare coupling constant $\hat \lambda$ is renormalized using 
\begin{align}\label{renl}
\frac{\hat{\lambda}}{\mu_0^\epsilon} S_\epsilon &= \frac{\lambda}{\mu_R^\epsilon} {Z}_{\lambda}(\mu_R^2)
\nonumber\\[1ex]
 &= \frac {\lambda}{\mu_R^\epsilon} \left[   1 + a_s \left( \frac{1}{\ep} r_{\lambda_{1;1}} \right)
 + a_s^2 \left( \frac{1}{\ep^2} r_{\lambda_{2;2}}
             + \frac{1}{\ep} r_{\lambda_{2;1}} \right) 
 + {\cal O}(a_s^3)  \right] \; ,
\end{align}
with $\lambda=\lambda(\mu_R^2)$ and
\begin{equation}
r_{\lambda_{1;1}} = 6 C_F \,, ~  
r_{\lambda_{2;2}} = \Big(18 C_F^2 + 6 \b0 C_F\Big)\, , ~  
r_{\lambda_{2,1}} =  \left( \frac{3}{2} C_F^2
                    + \frac{97}{6} C_F C_A
                    - \frac{10}{3} C_F T_F n_f \right).
\end{equation}
Using the eqn.(\ref{renas}) and eqn.(\ref{renl}), we now can express $|{\cal M } \rangle$ (eqn.(\ref{unm})) in powers  
of renormalized $a_s$ with UV finite matrix elements $|\rnM^{(i)} \rangle$  
\begin{equation} \label{renamp}
 |\rnM \rangle = \frac{\lambda}{\mu_R^\epsilon} \rnas^{\frac{1}{2}} \Bigg( |\rnM^{(0)} \rangle + a_s |\rnM^{(1)} \rangle + a_s^{2} |\rnM^{(2)} \rangle + {\cal O}({a}_s^3) \Bigg)
\end{equation}
where,
\begin{align} \label{rln}
 |\rnM^{(0)} \rangle &= \left(  \frac{1}{\mu^{\ep}_R} \right)^{\frac{1}{2}}    |\unM^{(0)} \rangle \; ,
\nonumber\\[1ex]
|\rnM^{(1)} \rangle &= \left(  \frac{1}{\mu^{\ep}_R} \right)^{\frac{3}{2}} 
\left[ ~ |\unM^{(1)} \rangle + \mu^{\ep}_R \Big( \frac{r_{a_1}}{2} + r_{\lambda_1} \Big) |\unM^{(0)} \rangle ~ \right] \; ,
\nonumber\\[1ex]
|\rnM^{(2)} \rangle &= \left( \frac{1}{\mu^{\ep}_R} \right)^{\frac{5}{2}} 
\Big[ ~ |\unM^{(2)} \rangle  
+ \mu^{\ep}_R \Big( \frac{3r_{a_1}}{2} + r_{\lambda_1} \Big)  |\unM^{(1)} \rangle 
\nonumber\\
& \qquad \qquad \quad
+ \mu^{2\ep}_R \left( \frac{r_{a_2}}{2} - \frac{r_{a_1}^2}{8} + \frac{r_{a_1}}{2} r_{\lambda_1}
+ r_{\lambda_2}  \right)  |\unM^{(0)} \rangle ~ \Big]
\end{align}
with
\begin{align}
 r_{a_1} &=  \left(\frac{1}{\ep} r_{a_{1;1}}\right)  \;, \quad \quad
 r_{a_2} =  \left(\frac{1}{\ep^2} r_{a_{2;2}} + \frac{1}{\ep} r_{a_{2;1}}\right)  \;,
\nonumber\\
 r_{\lambda_1} &=  \left(\frac{1}{\ep} r_{\lambda_{1;1}}\right)  \;, \quad \quad
 r_{\lambda_2} =  \left(\frac{1}{\ep^2} r_{\lambda_{2;2}} + \frac{1}{\ep} r_{\lambda_{2;1}} \right) \,.
\end{align}
We describe the 
computation of unrenormalized amplitudes $|\unM^{(l)} \rangle, l=0,1,2$ in
section \ref{sec:calc}. 
\subsection{Infrared factorization}
\label{sec:infrared}
In addition to UV divergences, the amplitude suffers from soft and collinear divergences
beyond leading order due to the presence of soft gluons and 
collinear massless partons in the loops.  
According to KLN theorem \cite{Kinoshita:1962ur, Lee:1964is},
to obtain infrared safe observables, we need to include appropriate contributions coming from real
emission processes along with mass factorization counter terms and to perform sum over
degenerate configurations.  Thanks to factorization properties of QCD amplitudes, the infrared
divergence structure of the amplitudes is well understood.  The earliest account on 
two loop QCD amplitudes was by Catani \cite{catani1}, who predicted the infrared poles in 
$\epsilon$ of multi-parton QCD amplitudes in dimensional regularization excluding two loop single 
pole.
In \cite{sterman}, Sterman and Tejeda-Yeomans 
demonstrated the connection of single pole in $\epsilon$ to a soft anomalous 
dimension matrix, later computed in \cite{Aybat:2006wq, Aybat:2006mz} using
factorization properties of the scattering amplitudes
along with infrared evolution equations.   
The decomposition of single pole term into universal collinear and soft
anomalous dimensions at two loop level in QCD was first observed 
in electromagnetic and Higgs form factors \cite{Ravindran:2004mb}.
Becher and Neubert \cite{Becher:2009cu}, using soft collinear effective theory, 
derived the exact formula for the infra-red divergences of 
scattering amplitudes with an arbitrary number of loops and legs
in massless QCD including single pole in dimensional regularization.  
Gardi and Magnea also arrived at, a similar all order result \cite{Gardi:2009qi} using
Wilson lines for hard partons and soft and eikonal jet functions in dimensional regularization.
Following Catani, we express the renormalized amplitudes $|{\cal M}^{(i)}\rangle$ 
in terms of the universal subtraction operators ${\bf I}^{(i)}_b(\epsilon)$ as
follows\footnote{\tiny{The numerical coefficients 2 and 4 with ${\bf I}^{(i)}$ come due to the different definition of $a_s$ between ours and Catani.}}

\begin{eqnarray}\label{catani1}
 |\rnM^{(1)}  \rangle &=& 2 \hspace{0.1cm} {\bf{I}}_{b}^{(1)} (\ep) \hspace{0.1cm} |\rnM^{(0)} \rangle
+ |\rnM^{(1)fin}  \rangle \, ,
\nonumber\\[2ex] 
|\rnM^{(2)} \rangle &=& 2 \hspace{0.1cm} {\bf{I}}_{b}^{(1)} (\ep) \hspace{0.1cm} |\rnM^{(1)} \rangle
+ 4 \hspace{0.1cm} {\bf{I}}^{(2)}_{b} (\ep) \hspace{0.1cm} |\rnM^{(0)}  \rangle
+ |\rnM^{(2)fin} \rangle 
\end{eqnarray} 
\begin{align}
\text{where,} ~~~~
 {\bf{I}}^{(1)}_{b} (\ep) &= \frac{1}{2}   \frac{e^{- \frac{\ep}{2} \gamma_E}}{\Gamma(1+\frac{\ep}{2})}
 \Bigg\{
 \Big( \frac{4}{\ep^2} - \frac{3}{\ep} \Big) (C_A - 2 C_F) \Big[ \Big( - \frac{s}{\mu^2_R} \Big)^{\frac{\ep}{2}} \Big]
\nonumber\\[1ex] 
& +
 \Big( - \frac{4 C_A}{\ep^2} + \frac{3 C_A}{ 2 \ep} + \frac{\b0}{2 \ep} \Big)
\left[ \Big( - \frac{t}{\mu^2_R} \Big)^{\frac{\ep}{2}} + \Big( - \frac{u}{\mu^2_R} \Big)^{\frac{\ep}{2}}   \right]
 \Bigg\} \; ,
\nonumber\\[1ex]
{\bf{I}}^{(2)}_{b} (\ep) &= - \hspace{0.1cm} \frac{1}{2} {\bf{I}}^{(1)}_{b} (\ep) \Big[ {\bf{I}}^{(1)}_{b} (\ep) - \frac{2 \b0}{\ep}  \Big]
 +\hspace{0.1cm} \frac{e^{\frac{\ep}{2} \gamma_E} \hspace{0.1cm} \Gamma(1+\ep)}{\Gamma(1+\frac{\ep}{2})} \Big[ -\frac{\b0}{\ep} + K \Big] \hspace{0.1cm} {\bf{I}}^{(1)}_{b} (2\ep) \hspace{1.4cm}
\nonumber\\[1ex]
& +\hspace{0.1cm} \left( 2 {\bf{H}}^{(2)}_{q} (\ep) + {\bf{H}}^{(2)}_{g} (\ep) \right)
\end{align}
with  
\begin{equation}
K = \left(\frac{67}{18}-\frac{\pi^2}{6} \right) C_A - \frac{10}{9} T_F n_f \, ,
\end{equation}

\begin{align}
 {\bf{H}}^{(2)}_q (\ep) &= \frac{1}{\ep}
 \Bigg\{
        C_A C_F \Big( - \frac{245}{432} + \frac{23}{16} \zeta_2 - \frac{13}{4} \zeta_3 \Big)
      + C_F^2 \Big( \frac{3}{16} - \frac{3}{2} \zeta_2 + 3 \zeta_3 \Big)
\nonumber\\[1ex]
& \quad
      + C_F n_f \Big( \frac{25}{216} - \frac{1}{8} \zeta_2 \Big) 
 \Bigg\} \; ,
\nonumber\\[1ex]
 {\bf{H}}^{(2)}_g (\ep) &= \frac{1}{\ep} 
\Bigg\{ 
        C_A^2 \Big( - \frac{5}{24} - \frac{11}{48} \zeta_2 - \frac{1}{4} \zeta_3 \Big) 
      + C_A n_f \Big(\frac{29}{54} +  \frac{1}{24} \zeta_2 \Big) - \frac{1}{4} C_F n_f -\frac{5}{54} n_f^2  
\Bigg\} \, .
\end{align}
The born amplitude $|\rnM^{(0)}\rangle$ and the finite parts $|\rnM^{(l)fin}\rangle, l=1,2$ are
 process dependent 
and hence they are determined by explicit computation. 
\section{Calculation of the amplitudes}
\label{sec:calc}
%% Figures 
%
\begin{figure}
 \centering 
 \includegraphics[width=0.8\textwidth]{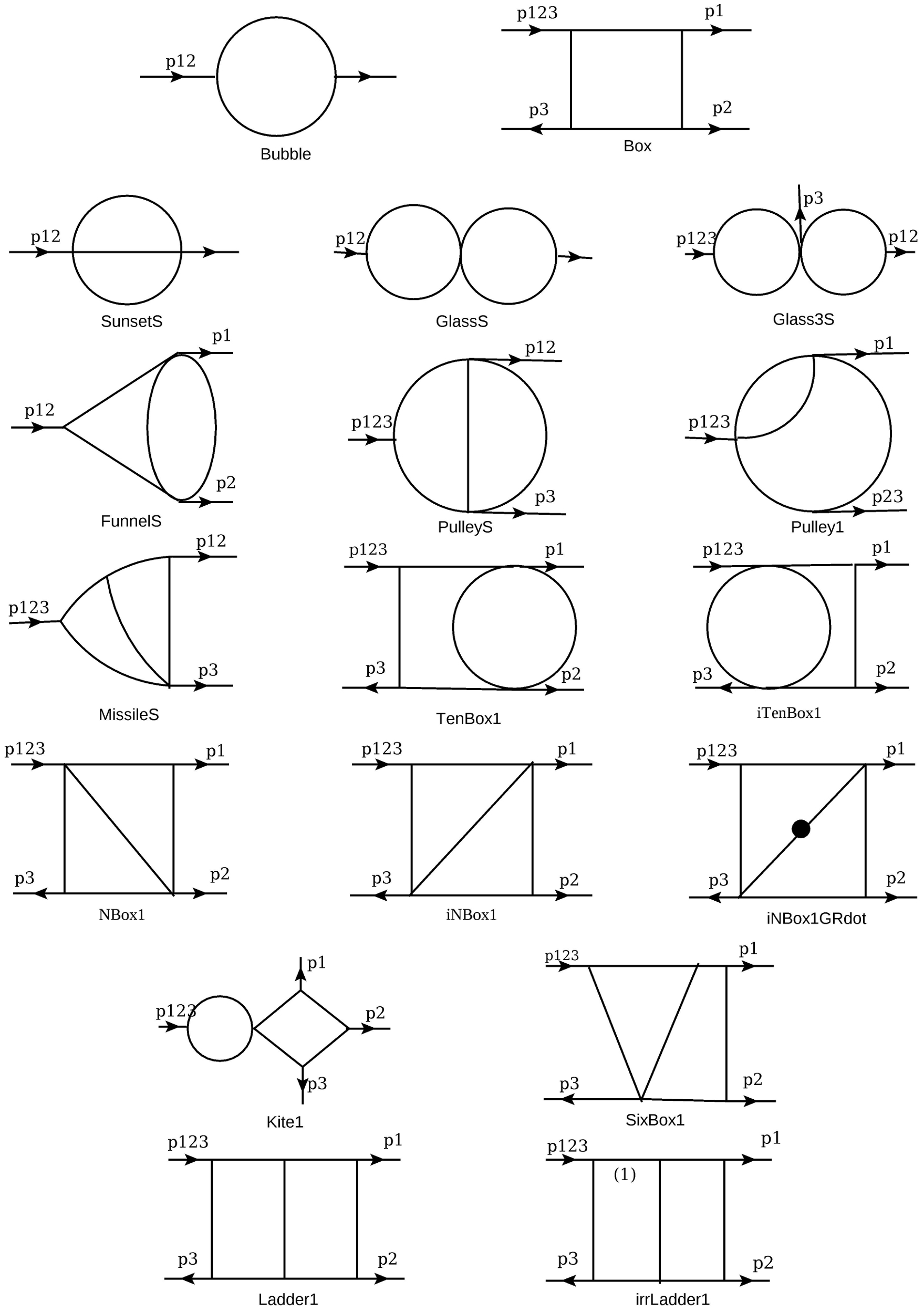}
\caption{Planar topologies of master integrals}
\label{fig:planar}
\end{figure}

\begin{figure}
\centering 
\includegraphics[width=0.88\textwidth]{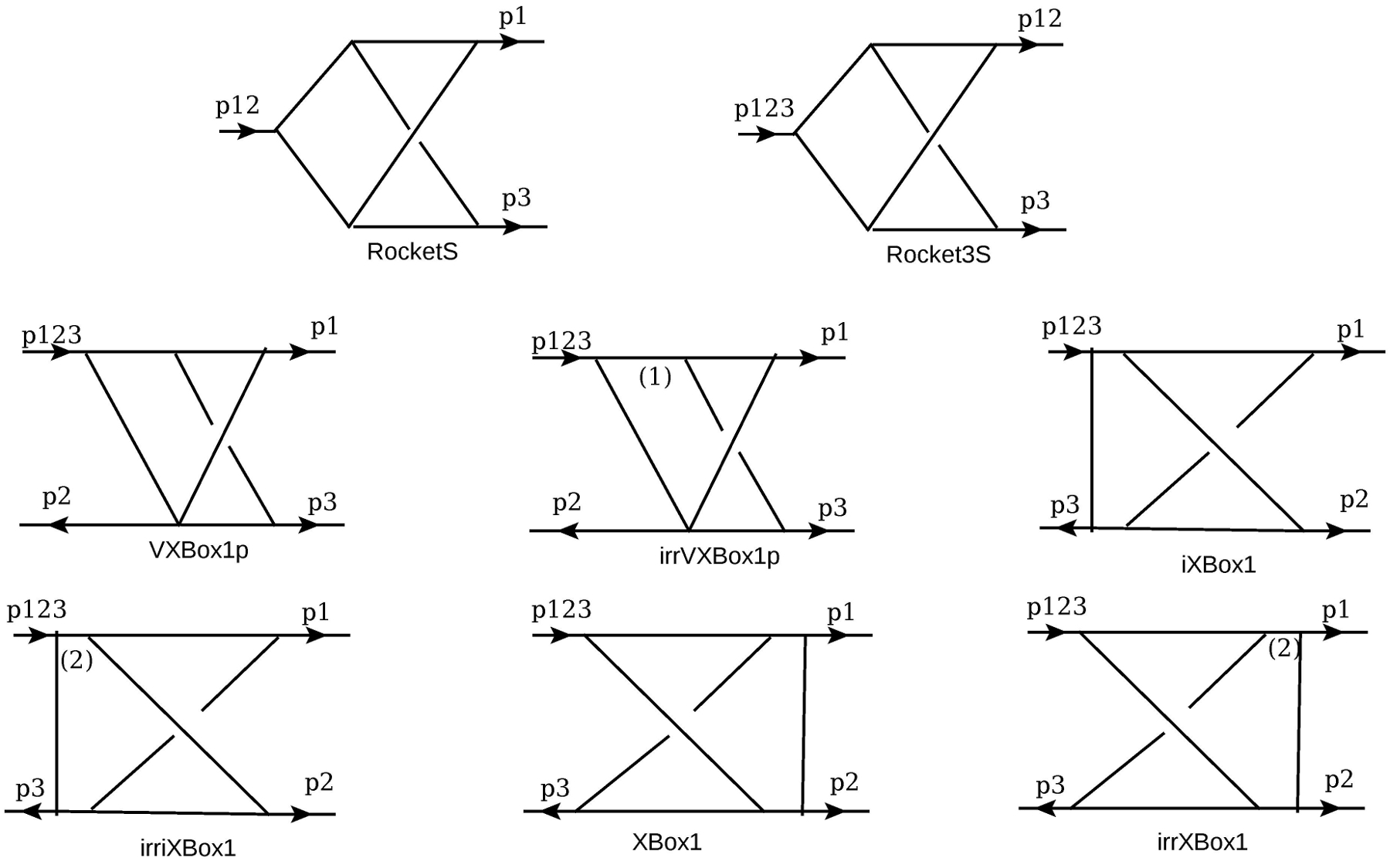}
\caption{Non-planar topologies of master integrals}
\label{fig:nonplanar}
\end{figure}
We now describe how we compute the coefficients $A_m$ from the amplitudes 
$|\hat {\cal M}^{(l)}\rangle$ for the
process  $H \rightarrow b+\overline b+g$ up to 
two loop level in QCD perturbation theory.      
QGRAF \cite{qgraf} is used to generate the Feynman amplitudes for this process.  
There are 2 diagrams at tree level, 13 at one loop and 251 at two loops excluding 
tadpole and self energy corrections to the external legs.  

Using FORM \cite{Vermaseren:2000nd} and Mathematica, output of the QGRAF is converted 
to a form suitable for further symbolic manipulation. 
Using the projectors given in eqn.(\ref{projectors}), we have projected out unrenormalized
$\hat{A}_i$ from these amplitudes.
They contain only scalar products among internal and external momenta.  
For the external on-shell gluon leg the physical polarization sum is done using
\begin{equation}
 \sum_s \varepsilon^{\mu}(p_3,s) \varepsilon^{\nu *}(p_3,s) = -g^{\mu\nu} + \frac{p_3^{\mu} q^{\nu} + q^{\mu} p_3^{\nu}}{p_3.q}
\end{equation}
where, $p_3$ is the gluon momentum and $q$ is an arbitrary light-like 4-vector
for which we choose $q = p_1$.
The Lorentz contractions and Dirac algebra are done 
in $d = 4 +\epsilon$ dimensions.  
The next step involves the evaluation of one and two loop tensor and scalar integrals.
This is done by first reducing them to an irreducible set of MIs using 
IBP identities and LI identities and 
substituting the MIs evaluated to desired accuracy in $\epsilon$.
We have used a Mathematica package
LiteRed \cite{litered} to use IBP \cite{chet} and LI identities \cite{gr} in an efficient manner.
The MIs for the kinematic configuration of the problem at hand 
are analytically known from the seminal works of Gehrmann and Remiddi \cite{Gehrmann:2000zt}.  
We use them to obtain the unrenormalized coefficients in a Laurent series in $\epsilon$.   
In order to optimize the use of LiteRed, we have 
reduced all the one and two loop integrals to belong to few integral sets.
This is done by shifting the loop momenta suitably using
an in-house algorithm which uses FORM. 
We find that the sets for both one and two loop integrals are exactly same as those given in \cite{Ahmed:2014gla}
for the case 
of massive spin-2 resonance $\rightarrow$ 3 gluons.
The topologies of the appearing planar and non-planar master integrals are shown
in fig.(\ref{fig:planar}) and fig.(\ref{fig:nonplanar}) respectively.
For one-loop diagrams, 
the integral belongs to one of the following sets:
\begin{eqnarray}\label{onebasis}
\{ \cD, \hspace{0.1cm} \cD_{1}, \hspace{0.1cm} \cD_{12}, \hspace{0.1cm} \cD_{123} \} \, ,
\{ \cD, \hspace{0.1cm} \cD_{2}, \hspace{0.1cm} \cD_{23}, \hspace{0.1cm} \cD_{123} \} \, ,\{ \cD, \hspace{0.1cm} \cD_{3}, \hspace{0.1cm} \cD_{31}, \hspace{0.1cm} \cD_{123} \}
\end{eqnarray}
where,
\begin{eqnarray}
 \cD = k_1^2, \hspace{0.1cm} \cD_{i} = (k_1 - p_i)^2,
\hspace{0.1cm} \cD_{ij} = (k_1 - p_i - p_j)^2,
\hspace{0.1cm} \cD_{ijk} = (k_1 - p_i - p_j - p_k)^2 \; .
\end{eqnarray}
At two loops, we have nine independent Lorentz invariants involving loop momenta $k_1$ and $k_2$, namely $\{ ( k_\alpha \cdot k_\beta ),
(k_\alpha \cdot p_i) \}, \alpha,\beta = 1, 2;\hspace{0.1cm} i = 1,...,3 $.  
Shifting of loop momenta allows us 
to express each two loop Feynman integral to contain terms 
belonging to one of the following six sets:  
\begin{eqnarray}\label{twobasis}
 &&\{ \cD_0, \hspace{0.1cm} \cD_1, \hspace{0.1cm} \cD_2, \hspace{0.1cm} \cD_{1;1}, \hspace{0.1cm} \cD_{2;1}, \hspace{0.1cm}
 \cD_{1;12}, \hspace{0.1cm} \cD_{2;12}, \hspace{0.1cm} \cD_{1;123}, \hspace{0.1cm} \cD_{2;123} \} \,,
\nonumber\\[1.5ex]
 &&\{ \cD_0, \hspace{0.1cm} \cD_1, \hspace{0.1cm} \cD_2, \hspace{0.1cm} \cD_{1;2}, \hspace{0.1cm} \cD_{2;2}, \hspace{0.1cm}
 \cD_{1;23}, \hspace{0.1cm} \cD_{2;23}, \hspace{0.1cm} \cD_{1;123}, \hspace{0.1cm} \cD_{2;123} \} \,,
\nonumber\\[1.5ex]
 &&\{ \cD_0, \hspace{0.1cm} \cD_1, \hspace{0.1cm} \cD_2, \hspace{0.1cm} \cD_{1;3}, \hspace{0.1cm} \cD_{2;3}, \hspace{0.1cm}
 \cD_{1;31}, \hspace{0.1cm} \cD_{2;31}, \hspace{0.1cm} \cD_{1;123}, \hspace{0.1cm} \cD_{2;123} \} \,,
\nonumber\\[1.5ex]
 &&\{ \cD_0, \hspace{0.1cm} \cD_1, \hspace{0.1cm} \cD_2, \hspace{0.1cm} \cD_{1;1}, \hspace{0.1cm} \cD_{2;1}, \hspace{0.1cm}
 \cD_{0;3}, \hspace{0.1cm} \cD_{1;12}, \hspace{0.1cm} \cD_{2;12}, \hspace{0.1cm} \cD_{1;123} \} \,,
\nonumber\\[1.5ex]
 &&\{ \cD_0, \hspace{0.1cm} \cD_1, \hspace{0.1cm} \cD_2, \hspace{0.1cm} \cD_{1;2}, \hspace{0.1cm} \cD_{2;2}, \hspace{0.1cm}
 \cD_{0;1}, \hspace{0.1cm} \cD_{1;23}, \hspace{0.1cm} \cD_{2;23}, \hspace{0.1cm} \cD_{1;123} \} \,,
\nonumber\\[1.5ex]
 &&\{ \cD_0, \hspace{0.1cm} \cD_1, \hspace{0.1cm} \cD_2, \hspace{0.1cm} \cD_{1;3}, \hspace{0.1cm} \cD_{2;3}, \hspace{0.1cm}
 \cD_{0;2}, \hspace{0.1cm} \cD_{1;31}, \hspace{0.1cm} \cD_{2;31}, \hspace{0.1cm} \cD_{1;123} \}
\end{eqnarray}
where,
\begin{eqnarray}
 && \cD_0 = (k_1 - k_2)^2, \hspace{0.1cm} \cD_{\alpha} = k_{\alpha}^2, \hspace{0.1cm} \cD_{\alpha;i} = (k_{\alpha} - p_i)^2,
\hspace{0.1cm} \cD_{\alpha; ij} = (k_{\alpha} - p_i - p_j)^2,
\nonumber\\[1ex]
&& \cD_{0;i} = (k_1 - k_2 - p_i)^2, \hspace{0.1cm} \cD_{\alpha;ijk} = (k_{\alpha} - p_i - p_j - p_k)^2 \; .
\end{eqnarray}
The UV singularities present in the bare coefficients are systematically removed using eqns.(\ref{renas} \& \ref{renl}).  
The resulting UV finite coefficients do contain divergences from soft and collinear partons.  
In the next section we will demonstrate that our results correctly reproduce 
divergences described in the section \ref{sec:infrared} at one and two loop level.  We will also 
present the finite parts of the coefficients $A_m$ up to two loop level.

\section{Results}
\label{sec:result}
In this section we present the results up to two loop level in QCD for the amplitude
$H \rightarrow b +\overline b+g$ in the ${\overline {MS}}$ scheme. 
The results are presented after subtracting the one and two loop universal subtraction operators
$I_b^{(i)}(\epsilon),i=1,2$ as described in the section \ref{sec:infrared}.
Following the eqns.(\ref{tencoeff}, \ref{coeffexpnd} \& \ref{renamp}), the 
$l^{th}$ loop amplitude can be written as
\begin{equation} \label{rnMxpnd}
 |\rnM^{(l)} \rangle = 4 \pi ~ T^a_{ij} \Big\{ A_1^{(l)} {\rm T_1} + A_2^{(l)} {\rm T_2} \Big\} \;.
\end{equation}
The renormalised coefficients $A_m^{(l)}$ are related to their bare counterparts $\hat A_m^{(l)}$ 
through
(see eqn.(\ref{rln})): 
\begin{align} \label{rln2}
 A_m^{(0)} &= \left(  \frac{1}{\mu^{\ep}_R} \right)^{\frac{1}{2}}    \hat{A}_m^{(0)} \; ,
\nonumber\\[1ex]
 A_m^{(1)} &= \left(  \frac{1}{\mu^{\ep}_R} \right)^{\frac{3}{2}} 
\left[ ~ \hat{A}_m^{(1)} + \mu^{\ep}_R \Big( \frac{r_{a_1}}{2} + r_{\lambda_1} \Big) \hat{A}_m^{(0)} ~ \right] \; ,
\nonumber\\[1ex]
 A_m^{(2)} &= \left( \frac{1}{\mu^{\ep}_R} \right)^{\frac{5}{2}} 
\Big[ ~ \hat{A}_m^{(2)}  
+ \mu^{\ep}_R \Big( \frac{3r_{a_1}}{2} + r_{\lambda_1} \Big)  \hat{A}_m^{(1)} 
\nonumber\\
& \qquad \qquad \quad
+ \mu^{2\ep}_R \left( \frac{r_{a_2}}{2} - \frac{r_{a_1}^2}{8} + \frac{r_{a_1}}{2} r_{\lambda_1}
+ r_{\lambda_2}  \right)  \hat{A}_m^{(0)} ~ \Big] \;.
\end{align}
Using the procedure discussed in the previous section, we first compute the bare 
coefficients $\hat A_m^{(l)}$ and the eqns.(\ref{rln2}) give the renormalized coefficients. 
The finite parts of the coefficients $A_m^{(l)}$ are defined 
after subtracting terms proportional to universal subtraction terms $I_b^{(l)}$ as follows 
\begin{align}\label{catani2}
 A_m^{(1)} &= 2 \hspace{0.1cm} {\bf{I}}_{b}^{(1)} (\ep) \hspace{0.1cm} A_m^{(0)}
+ A_m^{(1)fin} \,,
\nonumber\\[2ex] 
 A_m^{(2)} &= 2 \hspace{0.1cm} {\bf{I}}_{b}^{(1)} (\ep) \hspace{0.1cm} A_m^{(1)}
+ 4 \hspace{0.1cm} {\bf{I}}^{(2)}_{b} (\ep) \hspace{0.1cm} A_m^{(0)}
+ A_m^{(2)fin}
\end{align} 
where, we have used eqns.(\ref{tencoeff} \& \ref{catani1}). 

Expanding the right sides of eqns.(\ref{rln2} \& \ref{catani2}) in powers of $\ep$, we find that
the infrared poles agree exactly, providing a crucial test on the correctness of our
computation.  The finite parts of the coefficients have the following expansions:   
\begin{align}
 & A_m^{(l)fin} = \sum_{n=0}^l ~ A_m^{(0)} ~ {\cal B}^{(l)}_{m ; n} \ln^n \Big( - \frac{Q^2}{\mu^2} \Big) 
\end{align}
where,
\begin{equation}
 A_1^{(0)} = - \frac{4 i}{t ~ u} \hspace{1cm} \text{and} \hspace{1cm} A_2^{(0)} = i \Big( \frac{1}{t} + \frac{1}{u}  \Big)
\end{equation}
and the remaining coefficients ${\cal B}^{(l)}_{m;n}$ are given in the appendix.
We also performed an independent computation of 
$\langle {\cal M}^{(0)}|{\cal M}^{(l)}\rangle$ for $l=1,2$ without using any projectors
and then compared against one obtained using the projectors, i.e using
the coefficients $A_m^{(l)}$.  We find both give the same result, providing an independent
check on our computation.

Following \cite{Gehrmann:2002zr} \footnote{We thank T. Gehrmann for providing relevant analytically continued HPLs and 2d HPLs.}, 
we have obtained results for the crossed reactions given in eqn. (\ref{subprocess}) relevant for Higgs+1 jet production 
at hadron colliders.  The corresponding finite coefficients $A_m^{(l)fin}$ are attached with the arXiv submission.

\section{Conclusions}
\label{sec:conc}
\setcounter{equation}{0}
We have presented the amplitudes for the partonic subprocess
$H \rightarrow b+\overline b+g$ and other subprocesses related by crossing, up to two loop level in QCD 
that contribute to exclusive observables involving Higgs boson and a jet. 
The dominant one is from gluon gluon fusion which is already known to this accuracy.
We have used dimensional regularization to perform our computation.  Using appropriate projectors,
the amplitude is expressed in terms of two scalar coefficients $A_m$.  
We have found that the infrared structure of
the amplitude is according to Catani's prediction on QCD amplitudes upto two loop level.
Also, the coefficient of single pole term is found to be 
in agreement with predictions based on the observation
of the universal behavior of poles in the multi-parton QCD amplitudes. 

\section*{Acknowledgments} 

TA, MM and NR thank the Institute of Mathematical Sciences (IMSc) for the hospitality 
during the course of the work. We thank the staff of IMSc computer center
for their help. We sincerely thank T. Gehrmann for providing
us the master integrals and analytically continued HPLs and 2d-HPLs required for our computation.  We thank R. N. Lee for his help 
with LiteRed. Finally, we would like to thank K. Hasegawa, M. K. Mandal and L. Tancredi for useful discussions. 
The work of TA, MM and NR has been partially supported by funding from RECAPP,
Department of Atomic Energy, Govt. of India.

%%%%%%%%%%%%%%%%%%%%%%%%%%%%%%%%%%%%%%%%%%%%%%%%%%%%%%%%%%%%%%%%%%%%%%%%%%%%%%%%%%%%%%%%%%%

\begin{appendix}
\section{Harmonic polylogarithms}
\setcounter{equation}{0}

Here, we provide the definition of HPL and 2dHPL. HPL is represented by $H(\vec{m}_w;y)$ 
with a $w$-dimensional vector $\vec{m}_w$ of parameters and its argument $y$. The elements of $\vec{m}_w$ belong to $\{ 1, 0, -1 \}$ through which 
we define the following rational functions
\begin{equation}
 f(1;y) \equiv \frac{1}{1-y}, \qquad f(0;y) \equiv \frac{1}{y},  \qquad f(-1;y) \equiv \frac{1}{1+y} \, .
\end{equation}
The weight 1 $(w = 1)$ HPLs are 
\begin{equation}
 H(1, y) \equiv - \ln (1 - y), \qquad  H(0, y) \equiv \ln y, \qquad  H(-1, y) \equiv \ln (1 + y) \, .
\end{equation}
For $w > 1$, the definition of $H(m, \vec{m}_{w};y)$ is 
\begin{equation}\label{1dhpl}
 H(m, \vec{m}_w;y) \equiv \int_0^y dx ~ f(m, x) ~ H(\vec{m}_w;x),  \qquad \qquad  m \in 0, \pm 1  \, .
\end{equation}
The 2dHPLs are defined in the same way as eqn.(\ref{1dhpl}) with the new elements $\{ 2, 3 \}$ in $\vec{m}_w$ representing a new 
class of rational functions
\begin{equation}
 f(2;y) \equiv f(1-z;y) \equiv \frac{1}{1-y-z}, \qquad f(3;y) \equiv f(z;y) \equiv \frac{1}{y+z} 
\end{equation}
and correspondingly with the weight 1 $(w = 1)$ 2dHPLs
\begin{equation}
 H(2, y) \equiv - \ln \Big(1 - \frac{y}{1-z} \Big), \qquad  H(3, y) \equiv \ln \Big( \frac{y+z}{z} \Big) \, .
\end{equation}

\newpage
\section{One-loop coefficients} 
\setcounter{equation}{0}

{\tiny
\begin{dgroup*}
\begin{dmath*}
{\cal B}^{(1)}_{1 ; 1} = \frac{1}{6} (-11 C_A - 18 C_F + 2 n_f)  
\end{dmath*}
\begin{dmath*}
{\cal B}^{(1)}_{1 ; 0} =  
\frac{C_A}{6} \Big( -6 H(0,y) H(0,z)-6 H(0,y) H(1,z)-6 H(2,y) H(0,z)+12 H(3,y) \
H(1,z)-10 H(0,y)-9 H(2,y)-6 H(0,2,y)
-6 H(2,0,y)+12 H(3,2,y)-10 H(0,z)-9 \
H(1,z)+6 H(0,1,z)-6 H(1,0,z)- 6 \zeta_2 \Big)
+
C_F \Big( 2 H(0,y) H(1,z)- 4 H(3,y) H(1,z) 
+ 2 H(2,y) H(0,z)
+ 3 H(2,y)+12 \
H(0,2,y)- 2 H(1,0,y)+ 2 H(2,0,y)- 4 H(3,2,y)+ 3 H(1,z)- 2 H(0,1,z)- 2 \Big)
+
n_f \frac{1}{6} \Big( H(0,y)+H(0,z) \Big)
\end{dmath*}
\begin{dmath*}
{\cal B}^{(1)}_{2 ; 1} = \frac{1}{6} (-11 C_A-18 C_F + 2 n_f) 
\end{dmath*}
\begin{dmath*}
{\cal B}^{(1)}_{2 ; 0} = 
 \frac{C_A}{6} \Big( - 6 H(0,y) H(0,z)-6 H(0,y) H(1,z)-6 H(2,y) H(0,z)+12 H(3,y) \
H(1,z)-10 H(0,y)-9 H(2,y)-6 H(0,2,y)
-6 H(2,0,y)
+12 H(3,2,y)-10 H(0,z)-9 \
H(1,z)+6 H(0,1,z)-6 H(1,0,z) - 6 \zeta_2 + 6 \Big)
+
C_F \Big( 2 H(0,y) H(1,z)-4 H(3,y) H(1,z)
+2 H(2,y) H(0,z)
+3 H(2,y)+2 H(0,2,y)-2 \
H(1,0,y)+2 H(2,0,y)-4 H(3,2,y)+3 H(1,z)-2 H(0,1,z)-3 \Big)
+
n_f \frac{1}{6} \Big( H(0,y)+H(0,z) \Big)
\end{dmath*}
\end{dgroup*}
}

\section{Two-loop coefficients}
\setcounter{equation}{0}

{\tiny
\begin{dgroup*}
\begin{dmath*}
 {\cal B}^{(2)}_{1 ; 2} = \frac{1}{24} \Big( 121 C_A^2 + 44 C_A  ( 6 C_F - n_f)
+ 4 \Big( 27 C_F^2 - 12 C_F n_f + n_f^2 \Big) \Big)  
\end{dmath*}
\begin{dmath*}
 {\cal B}^{(2)}_{1 ; 1} = 
C_A^2 \frac{1}{108} \Big( 594 H(0,y) H(0,z)+594 H(0,y) H(1,z)+594 H(2,y) H(0,z)-1188 \
H(3,y) H(1,z)+990 H(0,y)+891 H(2,y)
+594 H(0,2,y)+594 H(2,0,y)-1188 \
H(3,2,y)+990 H(0,z)+891 H(1,z)-594 H(0,1,z)+594 H(1,0,z)+495 \
\zeta_2-108 \zeta_3 -702 \Big)
+
C_A C_F  \frac{1}{108} \Big( -1188 H(0,y) H(1,z)+54 H(0,y) (6 H(0,z)+6 H(1,z)+10)-1188 \
H(2,y) H(0,z)+54 (6 H(2,y)+10) H(0,z)+1728 H(3,y) H(1,z)-1296 H(2,y)-864 \
H(0,2,y)+1188 H(1,0,y)-864 H(2,0,y)+1728 H(3,2,y)-1296 H(1,z)+864 \
H(0,1,z)+324 H(1,0,z)+1566 \zeta_2-2808 \zeta_3-1048 \Big)
+
C_F^2 \frac{1}{108} \Big( -648 H(0,y) H(1,z)-648 H(2,y) H(0,z)
\end{dmath*}
\begin{dmath*}
{\white=}
+1296 H(3,y) H(1,z)-972 \
H(2,y)-648 H(0,2,y)+648 H(1,0,y)-648 H(2,0,y)+1296 H(3,2,y)-972 H(1,z)+648 \
H(0,1,z)-1296 \zeta_2 + 2592 \zeta_3 + 648 \Big)
+
C_A n_f \frac{1}{108} \Big( - 2 \Big(-27 (4 H(3,y)-3) H(1,z)+81 H(2,y)+54 H(0,2,y)+54 \
H(2,0,y)-108 H(3,2,y)-54 H(0,1,z)+54 H(1,0,z) + 45 \zeta_2 - 206 \Big) -9 H(0,y) \
(12 H(0,z)+12 H(1,z)+31)-9 (12 H(2,y)+31) H(0,z) \Big)
+
C_F n_f \frac{1}{54} \Big( 108 H(0,y) H(1,z)+108 H(2,y) H(0,z)-216 H(3,y) H(1,z)-27 H(0,y)+162 \
H(2,y)+108 H(0,2,y)-108 H(1,0,y)+108 H(2,0,y)-216 H(3,2,y)-27 H(0,z)+162 \
H(1,z)-108 H(0,1,z) - 54 \zeta_2 + 32 \Big)
+
n_f^2 \frac{1}{54} \Big( 9 H(0,y)+9 H(0,z)-20 \Big)
\end{dmath*}
% \end{dgroup*}
% 
% \begin{dgroup*}
\begin{dmath*}
 {\cal B}^{(2)}_{1 ; 0} = 
C_A^2  \Bigg\{
\zeta_4 \Big( 39/8 \Big)
+
\zeta_3 \Bigg(
- H(0,y) - 6 H(1,y) + 11 H(2,y) - H(0,z) + 5 H(1,z) + 6 \tpu/s + 407/36
\Bigg)
+
\zeta_2 \Bigg(
108 s^2 H(0,y) H(0,z)
\end{dmath*}
\begin{dmath*}
{\white=}
+72 s^2 H(1,y) H(0,z)+108 s^2 H(0,y) H(1,z)+72 s^2 \
H(1,y) H(1,z)+36 s^2 H(2,y) H(0,z)-72 s^2 H(3,y) H(1,z)+147 s^2 H(0,y)+453 \
s^2 H(2,y)+108 s^2 H(0,2,y)+72 s^2 H(1,2,y)+36 s^2 H(2,0,y)-72 s^2 \
H(3,2,y)+147 s^2 H(0,z)+237 s^2 H(1,z)+36 s^2 H(0,1,z)+36 s^2 H(1,0,z)+72 \
s^2 H(1,1,z)-216 s \spt H(1,y)-108 s t H(0,z)-108 s u H(0,y)-216 s u \
H(1,z)-40 s^2+96 t u
\Bigg) \Big/ (36 s^2)
+
\Bigg(
-\frac{13}{4} H(2,y) H(1,z)+\frac{134}{9} H(3,y) H(1,z)+7 H(0,0,y) \
H(1,z)-\frac{2}{3} H(0,2,y) H(1,z)-7 H(0,3,y) H(1,z)-\frac{2}{3} H(2,0,y) \
H(1,z)+\frac{13}{3} H(2,3,y) H(1,z)-7 H(3,0,y) H(1,z)+\frac{4}{3} H(3,2,y) \
H(1,z)+\frac{22}{3} H(3,3,y) H(1,z)+2 H(0,0,2,y) H(1,z)+2 H(0,2,0,y) \
H(1,z)+4 H(0,3,0,y) H(1,z)+4 H(0,3,3,y) H(1,z)+2 H(1,0,3,y) H(1,z)-2 \
H(1,2,3,y) H(1,z)+2 H(2,0,0,y) H(1,z)+2 H(2,0,3,y) H(1,z)+2 H(2,1,0,y) \
H(1,z)+4 H(2,2,3,y) H(1,z)+2 H(2,3,0,y) H(1,z)-8 H(2,3,3,y) H(1,z)+4 \
H(3,0,3,y) H(1,z)+4 H(3,3,0,y) H(1,z)-16 H(3,3,3,y) H(1,z)
\end{dmath*}
\begin{dmath*}
{\white=}
-\frac{361}{54} \
H(1,z)-\frac{361}{54} H(2,y)+\frac{80}{9} H(0,0,y)+7 H(2,y) H(0,0,z)+2 \
H(0,0,y) H(0,0,z)+\frac{80}{9} H(0,0,z)+\frac{11}{3} H(2,y) \
H(0,1,z)+\frac{1}{3} H(3,y) H(0,1,z)+2 H(0,0,y) H(0,1,z)+\frac{179}{18} \
H(0,1,z)+2 H(0,0,z) H(0,2,y)-2 H(0,1,z) H(0,2,y)-\frac{89}{18} H(0,2,y)-2 \
H(0,1,z) H(0,3,y)+\frac{8 t u H(1,0,y)}{3 s^2}+\frac{8 t u H(1,0,z)}{3 \
s^2}+6 H(2,y) H(1,0,z)-7 H(3,y) H(1,0,z)+2 H(0,0,y) H(1,0,z)-6 H(0,3,y) \
H(1,0,z)-\frac{89}{18} H(1,0,z)+\frac{4}{3} H(3,y) H(1,1,z)+2 H(0,0,y) \
H(1,1,z)-\frac{13}{4} H(1,1,z)-2 H(0,1,z) H(1,2,y)
{\white \Bigg\}}
\end{dmath*}
\end{dgroup*}
\begin{dgroup*}
\begin{dmath*}
{\white=}
{\white \Bigg\{}
+2 H(1,0,z) H(1,2,y)+2 \
H(0,0,z) H(2,0,y)+2 H(0,1,z) H(2,0,y)+2 H(1,0,z) H(2,0,y)-\frac{89}{18} \
H(2,0,y)+2 H(0,0,z) H(2,2,y)+4 H(0,1,z) H(2,2,y)+4 H(1,0,z) \
H(2,2,y)-\frac{13}{4} H(2,2,y)-6 H(0,1,z) H(2,3,y)+2 H(1,0,z) H(2,3,y)+2 \
H(0,1,z) H(3,0,y)-2 H(1,0,z) H(3,0,y)+\frac{134}{9} H(3,2,y)-12 H(0,1,z) \
H(3,3,y)+4 H(1,0,z) H(3,3,y)-2 H(1,y) H(0,0,1,z)-2 H(2,y) H(0,0,1,z)-8 \
H(3,y) H(0,0,1,z)+\frac{1}{3} H(0,0,1,z)+7 H(0,0,2,y)+\frac{3 u \
H(0,1,0,y)}{s}+\frac{3 t H(0,1,0,z)}{s}+2 H(1,y) H(0,1,0,z)+6 H(2,y) \
H(0,1,0,z)+4 H(3,y) H(0,1,0,z)+\frac{2}{3} H(0,1,1,z)+7 \
H(0,2,0,y)-\frac{2}{3} H(0,2,2,y)-7 H(0,3,2,y)+2 H(2,y) H(1,0,0,z)+7 \
H(1,0,0,z)-2 H(1,y) H(1,0,1,z)+6 H(2,y) H(1,0,1,z)+\frac{11}{3} \
H(1,0,1,z)-\frac{6 \spt H(1,1,0,y)}{s}-\frac{6 u H(1,1,0,z)}{s}+2 \
H(1,y) H(1,1,0,z)+6 H(2,y) H(1,1,0,z)+H(0,y) \Big(-\frac{3 H(1,0,z) \
u}{s}+\frac{8 u}{3 s}+\Big(\frac{8 t u}{3 s^2}-\frac{14}{3}\Big) \
H(0,z)-\frac{89}{18} H(1,z)+7 H(0,0,z)-\frac{13}{6} H(0,1,z)+\frac{17}{2} \
H(1,0,z)-\frac{2}{3} H(1,1,z)-2 H(0,0,1,z)-2 H(0,1,1,z)+2 H(1,0,0,z)+2 \
H(1,1,0,z)-\frac{142}{27}\Big)+7 H(2,0,0,y)-\frac{2}{3} \
H(2,0,2,y)+\frac{20}{3} H(2,1,0,y)-\frac{2}{3} H(2,2,0,y)+\frac{13}{3} \
H(2,3,2,y)-7 H(3,0,2,y)-7 H(3,2,0,y)+\frac{4}{3} H(3,2,2,y)+\frac{22}{3} \
H(3,3,2,y)+H(0,z) \Big(\frac{8 t}{3 s}-\frac{89}{18} H(2,y)+7 \
H(0,0,y)+\frac{29}{6} H(0,2,y)-\frac{3 \spt H(1,0,y)}{s}+\frac{23}{2} \
H(2,0,y)-\frac{2}{3} H(2,2,y)-7 H(3,2,y)+2 H(0,0,2,y)+2 H(0,2,0,y)-2 \
H(0,2,2,y)-6 H(0,3,2,y)-2 H(1,0,2,y)+2 H(2,0,0,y)+2 H(2,2,0,y)+2 \
H(2,3,2,y)-2 H(3,0,2,y)-2 H(3,2,0,y)+4 H(3,3,2,y)-\frac{142}{27}\Big)+8 \
H(0,0,1,0,z)+2 H(0,0,1,1,z)+2 H(0,0,2,2,y)+2 H(0,1,0,1,z)+6 H(0,1,1,0,z)+2 \
H(0,2,0,2,y)
\end{dmath*}
\begin{dmath*}
{\white=}
+4 H(0,2,1,0,y)+2 H(0,2,2,0,y)+4 H(0,3,0,2,y)+4 H(0,3,2,0,y)+4 \
H(0,3,3,2,y)+6 H(1,0,1,0,z)+2 H(1,0,3,2,y)+2 H(1,1,0,0,z)+4 H(1,1,0,1,z)+6 \
H(1,1,1,0,z)-2 H(1,2,3,2,y)+2 H(2,0,0,2,y)+2 H(2,0,1,0,y)+2 H(2,0,2,0,y)+2 \
H(2,0,3,2,y)+2 H(2,1,0,2,y)+2 H(2,1,2,0,y)+2 H(2,2,0,0,y)+4 H(2,2,1,0,y)+4 \
H(2,2,3,2,y)+2 H(2,3,0,2,y)+2 H(2,3,2,0,y)-8 H(2,3,3,2,y)+4 H(3,0,3,2,y)+4 \
H(3,3,0,2,y)+4 H(3,3,2,0,y)-16 H(3,3,3,2,y)-\frac{571}{81}
\Bigg)
\Bigg\}
\end{dmath*}
\begin{dmath*}
{\white=}
+
C_A C_F  \Bigg\{
\zeta_4 \Big( \frac{93}{4} \Big)
+
\zeta_3 \Big(
22 H(1,y)-18 H(2,y)+4 H(1,z)+\frac{4 t^2}{s^2}+\frac{4 u^2}{s^2}-\frac{8 \
\tpu}{s}+\frac{335}{18}
\Big)
+
\zeta_2 \Big(
-\frac{4 t^2 H(1,y)}{s^2}-\frac{2 t^2 H(0,z)}{s^2}-\frac{2 u^2 \
H(0,y)}{s^2}-\frac{4 u^2 H(1,z)}{s^2}+\frac{8 t H(1,y)}{s}+\frac{4 t \
H(0,z)}{s}+\frac{4 u H(0,y)}{s}+\frac{8 u H(1,z)}{s}-\frac{107 t^2 H(2,y)}{6 \
{\tpu}^2}-\frac{79 t^2 H(1,z)}{6 {\tpu}^2}-\frac{119 t u H(2,y)}{3 \
{\tpu}^2}-\frac{91 t u H(1,z)}{3 {\tpu}^2}-\frac{107 u^2 H(2,y)}{6 \
{\tpu}^2}-\frac{79 u^2 H(1,z)}{6 {\tpu}^2}-2 H(1,y) H(0,z)-6 H(0,y) \
H(1,z)-2 H(1,y) H(1,z)-4 H(2,y) H(0,z)+\frac{14}{3} H(1,y)+2 H(0,2,y)+6 \
H(1,0,y)-2 H(1,2,y)-4 H(2,0,y)-8 H(2,1,y)+8 H(2,2,y)+2 H(0,1,z)+2 H(1,0,z)-2 \
H(1,1,z)-\frac{18 t u}{s^2}-\frac{2 t^2}{s {\tpu}}-\frac{2 u^2}{s \
{\tpu}}-\frac{17}{6}
\Big)
+
\Bigg(
-\frac{2 H(1,0,y) t^2}{s {\tpu}}-\frac{5 H(2,y) H(1,0,z) t^2}{3 \
{\tpu}^2}+\frac{2 H(1,0,z) t^2}{s {\tpu}}+\frac{10 H(0,1,0,y) t^2}{3 \
{\tpu}^2}
\end{dmath*}
\begin{dmath*}
{\white=}
+\frac{2 H(0,1,0,z) t^2}{s^2}+\frac{28 H(0,1,0,z) t^2}{3 \
{\tpu}^2}-\frac{4 H(1,1,0,y) t^2}{s^2}+\frac{6 H(1,1,0,z) \
t^2}{{\tpu}^2}+\frac{3 H(2,1,0,y) t^2}{{\tpu}^2}-\frac{3 H(1,z) \
H(3,y) t}{u}+\frac{4 u H(1,0,y) t}{s {\tpu}}-\frac{18 u H(1,0,y) \
t}{s^2}-\frac{152 H(1,0,y) t}{9 {\tpu}}+\frac{4 u H(1,0,z) t}{s \
{\tpu}}-\frac{18 u H(1,0,z) t}{s^2}-\frac{22 u H(2,y) H(1,0,z) t}{3 \
{\tpu}^2}-\frac{H(1,0,z) t}{{\tpu}}-\frac{3 H(3,2,y) t}{u}+\frac{32 \
u H(0,1,0,y) t}{3 {\tpu}^2}+\frac{68 u H(0,1,0,z) t}{3 \
{\tpu}^2}-\frac{4 H(0,1,0,z) t}{s}+\frac{8 H(1,1,0,y) t}{s}+\frac{8 u \
H(1,1,0,z) t}{{\tpu}^2}+\frac{2 u H(2,1,0,y) \
t}{{\tpu}^2}+\frac{604}{27} H(1,z)+2 H(1,z) H(2,y)+\frac{604}{27} \
H(2,y)-\frac{304}{9} H(1,z) H(3,y)-14 H(1,z) H(0,0,y)-14 H(2,y) \
H(0,0,z)-\frac{22}{3} H(2,y) H(0,1,z)-\frac{20}{3} H(3,y) H(0,1,z)-4 \
H(0,0,y) H(0,1,z)-\frac{197}{9} H(0,1,z)-\frac{14}{3} H(1,z) H(0,2,y)-4 \
H(0,0,z) H(0,2,y)+6 H(0,1,z) H(0,2,y)+\frac{107}{9} H(0,2,y)+20 H(1,z) \
H(0,3,y)+4 H(0,1,z) H(0,3,y)+\frac{2 u^2 H(1,0,y)}{s {\tpu}}-\frac{143 u \
H(1,0,y)}{9 {\tpu}}+3 H(1,z) H(1,0,y)
\end{dmath*}
\begin{dmath*}
{\white=}
-2 H(0,1,z) H(1,0,y)-\frac{2 u^2 \
H(1,0,z)}{s {\tpu}}-\frac{5 u H(1,0,z)}{{\tpu}}-\frac{5 u^2 H(2,y) \
H(1,0,z)}{3 {\tpu}^2}+20 H(3,y) H(1,0,z)-4 H(0,0,y) H(1,0,z)+8 H(0,2,y) \
H(1,0,z)+16 H(0,3,y) H(1,0,z)+2 H(1,0,y) H(1,0,z)+\frac{28}{3} H(3,y) \
H(1,1,z)-8 H(0,0,y) H(1,1,z)+4 H(0,3,y) H(1,1,z)+2 H(1,1,z)+6 H(0,1,z) \
H(1,2,y)-2 H(1,0,z) H(1,2,y)-\frac{14}{3} H(1,z) H(2,0,y)-4 H(0,0,z) \
H(2,0,y)-2 H(0,1,z) H(2,0,y)-6 H(1,0,z) H(2,0,y)+\frac{107}{9} H(2,0,y)-8 \
H(0,0,z) H(2,2,y)-16 H(0,1,z) H(2,2,y)-8 H(1,0,z) H(2,2,y)+2 \
H(2,2,y)-\frac{8}{3} H(1,z) H(2,3,y)+12 H(0,1,z) H(2,3,y)-4 H(1,0,z) \
H(2,3,y)+20 H(1,z) H(3,0,y)-4 H(0,1,z) H(3,0,y)+4 H(1,1,z) \
H(3,0,y)+\frac{28}{3} H(1,z) H(3,2,y)-4 H(0,1,z) H(3,2,y)-4 H(1,0,z) \
H(3,2,y)-\frac{304}{9} H(3,2,y)-\frac{80}{3} H(1,z) H(3,3,y)+28 H(0,1,z) \
H(3,3,y)
-12 H(1,0,z) H(3,3,y)-8 H(1,1,z) H(3,3,y)+6 H(1,y) H(0,0,1,z)+16 \
H(3,y) H(0,0,1,z)-\frac{2}{3} H(0,0,1,z)
-8 H(1,z) H(0,0,2,y)-14 H(0,0,2,y)+4 \
H(1,z) H(0,0,3,y)+\frac{2 u^2 H(0,1,0,y)}{s^2}+\frac{10 u^2 H(0,1,0,y)}{3 \
{\tpu}^2}-\frac{4 u H(0,1,0,y)}{s}+2 H(1,z) H(0,1,0,y)+\frac{28 u^2 \
H(0,1,0,z)}{3 {\tpu}^2}-2 H(1,y) H(0,1,0,z)-12 H(2,y) H(0,1,0,z)-4 \
H(3,y) H(0,1,0,z)
\end{dmath*}
\begin{dmath*}
{\white=}
-4 H(3,y) H(0,1,1,z)+\frac{14}{3} H(0,1,1,z)-8 H(1,z) \
H(0,2,0,y)-14 H(0,2,0,y)-\frac{14}{3} H(0,2,2,y)+2 H(1,z) H(0,2,3,y)-8 \
H(1,z) H(0,3,0,y)+4 H(1,z) H(0,3,2,y)+20 H(0,3,2,y)-12 H(1,z) H(0,3,3,y)+4 \
H(1,z) H(1,0,0,y)+14 H(1,0,0,y)-4 H(2,y) H(1,0,0,z)+6 H(1,y) H(1,0,1,z)-18 \
H(2,y) H(1,0,1,z)-4 H(3,y) H(1,0,1,z)-\frac{13}{3} H(1,0,1,z)+3 H(1,0,2,y)-6 \
H(1,z) H(1,0,3,y)+\frac{14}{3} H(1,1,0,y)+H(0,y) \Big(\Big(-\frac{18 t \
u}{s^2}+\frac{4 t u}{s {\tpu}}+2\Big) H(0,z)+\Big(\frac{3 \
u}{t}+\frac{107}{9}\Big) H(1,z)+\frac{1}{3} \Big(-\frac{19 H(1,0,z) \
t^2}{{\tpu}^2}-\frac{54 u t}{s {\spt}}-\frac{50 u H(1,0,z) \
t}{{\tpu}^2}+\frac{10 t}{{\spt}}-\frac{36 u}{{\spt}}+13 \
H(0,1,z)-\frac{6 u^2 H(1,0,z)}{s^2}-\frac{19 u^2 \
H(1,0,z)}{{\tpu}^2}+\frac{12 u H(1,0,z)}{s}-14 H(1,1,z)+12 H(0,0,1,z)+6 \
H(0,1,0,z)+24 H(0,1,1,z)+6 H(1,0,1,z)-12 H(1,1,0,z)+\frac{10 \
s}{{\spt}}\Big)\Big)-\frac{4 u^2 H(1,1,0,z)}{s^2}+\frac{6 u^2 \
H(1,1,0,z)}{{\tpu}^2}+\frac{8 u H(1,1,0,z)}{s}-2 H(1,y) H(1,1,0,z)-18 \
H(2,y) H(1,1,0,z)-4 H(3,y) H(1,1,0,z)+3 H(1,2,0,y)+6 H(1,z) H(1,2,3,y)-8 \
H(1,z) H(2,0,0,y)-14 H(2,0,0,y)-\frac{14}{3} H(2,0,2,y)-4 H(1,z) \
H(2,0,3,y)
\end{dmath*}
\begin{dmath*}
{\white=}
+\frac{3 u^2 H(2,1,0,y)}{{\tpu}^2}-2 H(1,z) \
H(2,1,0,y)-\frac{14}{3} H(2,2,0,y)-16 H(1,z) H(2,2,3,y)-4 H(1,z) \
H(2,3,0,y)-\frac{8}{3} H(2,3,2,y)+16 H(1,z) H(2,3,3,y)+4 H(1,z) \
H(3,0,2,y)+20 H(3,0,2,y)-12 H(1,z) H(3,0,3,y)+4 H(1,z) H(3,2,0,y)+20 \
H(3,2,0,y)+\frac{28}{3} H(3,2,2,y)-8 H(1,z) H(3,2,3,y)-12 H(1,z) \
H(3,3,0,y)+H(0,z) \Big(\Big(\frac{3 t}{u}+\frac{107}{9}\Big) \
H(2,y)+\frac{1}{3} \Big(-\frac{47 H(2,0,y) t^2}{{\tpu}^2}-\frac{54 u \
t}{s {\spu}}-\frac{106 u H(2,0,y) t}{{\tpu}^2}-\frac{36 \
t}{{\spu}}+\frac{10 u}{{\spu}}-47 H(0,2,y)+\Big(-\frac{6 \
t^2}{s^2}+\frac{12 t}{s}+28\Big) H(1,0,y)-\frac{47 u^2 \
H(2,0,y)}{{\tpu}^2}-14 H(2,2,y)+60 H(3,2,y)-24 H(0,0,2,y)+6 \
H(0,1,0,y)+12 H(0,2,2,y)+48 H(0,3,2,y)+12 H(1,0,0,y)+12 H(1,0,2,y)+6 \
H(1,2,0,y)-12 H(2,0,0,y)+6 H(2,0,2,y)-6 H(2,1,0,y)-12 H(2,2,0,y)-12 \
H(2,3,2,y)+24 H(3,0,2,y)+12 H(3,2,2,y)-36 H(3,3,2,y)+\frac{10 \
s}{{\spu}}\Big)\Big)-8 H(1,z) H(3,3,2,y)-\frac{80}{3} H(3,3,2,y)+40 \
H(1,z) H(3,3,3,y)+4 H(0,0,1,0,y)-8 H(0,0,1,0,z)-8 H(0,0,1,1,z)-8 \
H(0,0,2,2,y)+4 H(0,0,3,2,y)-8 H(0,1,0,1,z)+2 H(0,1,0,2,y)-12 H(0,1,1,0,z)+2 \
H(0,1,2,0,y)-8 H(0,2,0,2,y)-2 H(0,2,1,0,y)-8 H(0,2,2,0,y)+2 H(0,2,3,2,y)-8 \
H(0,3,0,2,y)-8 H(0,3,2,0,y)+4 H(0,3,2,2,y)-12 H(0,3,3,2,y)-2 H(1,0,0,1,z)+4 \
H(1,0,0,2,y)+4 H(1,0,1,0,y)-4 H(1,0,1,0,z)+4 H(1,0,2,0,y)-6 H(1,0,3,2,y)-12 \
H(1,1,0,1,z)-14 H(1,1,1,0,z)+4 H(1,2,0,0,y)+4 H(1,2,1,0,y)+6 H(1,2,3,2,y)-8 \
H(2,0,0,2,y)-8 H(2,0,2,0,y)-4 H(2,0,3,2,y)+4 H(2,1,0,0,y)-2 H(2,1,0,2,y)-8 \
H(2,1,1,0,y)-2 H(2,1,2,0,y)-8 H(2,2,0,0,y)-8 H(2,2,1,0,y)-16 H(2,2,3,2,y)-4 \
H(2,3,0,2,y)-4 H(2,3,2,0,y)+16 H(2,3,3,2,y)+8 H(3,0,1,0,y)+4 H(3,0,2,2,y)-12 \
H(3,0,3,2,y)+4 H(3,2,0,2,y)-8 H(3,2,1,0,y)+4 H(3,2,2,0,y)-8 H(3,2,3,2,y)-12 \
H(3,3,0,2,y)-12 H(3,3,2,0,y)-8 H(3,3,2,2,y)+40 \
H(3,3,3,2,y)-\frac{467}{81}-\frac{3 u H(1,z) H(3,y)}{t}-\frac{3 u \
H(0,1,z)}{t}+\frac{3 u H(0,2,y)}{t}-\frac{3 u^2 H(1,0,y)}{{\tpu} \
t}+\frac{3 u H(2,0,y)}{t}-\frac{3 u H(3,2,y)}{t}
\Bigg)
\Bigg\}
\end{dmath*}
\begin{dmath*}
{\white=}
+
C_F^2  \Bigg\{
\zeta_4 (-22)
+
\zeta_3 \Big(
-16 H(1,y)+8 H(2,y)-8 H(1,z)-\frac{4 t^2}{s^2}-\frac{4 u^2}{s^2}+\frac{8 \
\tpu}{s}-30
\Big)
+
\zeta_2 \Big(
\frac{4 t^2 H(1,y)}{s^2}+\frac{2 t^2 H(0,z)}{s^2}+\frac{2 u^2 \
H(0,y)}{s^2}+\frac{4 u^2 H(1,z)}{s^2}-\frac{8 t H(1,y)}{s}-\frac{4 t \
H(0,z)}{s}-\frac{4 u H(0,y)}{s}-\frac{8 u H(1,z)}{s}+\frac{6 t^2 \
H(2,y)}{{\tpu}^2}+\frac{6 t^2 H(1,z)}{{\tpu}^2}+\frac{16 t u \
H(2,y)}{{\tpu}^2}+\frac{16 t u H(1,z)}{{\tpu}^2}+\frac{6 u^2 \
H(2,y)}{{\tpu}^2}+\frac{6 u^2 H(1,z)}{{\tpu}^2}-8 H(2,y) H(1,z)+8 \
H(0,1,y)-8 H(0,2,y)+8 H(1,1,y)+8 H(2,1,y)-16 H(2,2,y)+\frac{12 t \
u}{s^2}+\frac{2 t^2}{s {\tpu}}+\frac{2 u^2}{s {\tpu}}+16
\Big)
+ \Bigg(
\frac{2 H(1,0,y) t^2}{s {\tpu}}-\frac{2 H(1,0,z) t^2}{s \
{\tpu}}+\frac{12 H(0,1,0,y) t^2}{{\tpu}^2}-\frac{2 H(0,1,0,z) \
t^2}{s^2}+\frac{4 H(1,1,0,y) t^2}{s^2}-\frac{6 H(1,1,0,z) \
t^2}{{\tpu}^2}-\frac{12 H(2,1,0,y) t^2}{{\tpu}^2}+\frac{6 H(1,z) \
H(3,y) t}{u}-\frac{4 u H(1,0,y) t}{s {\tpu}}+\frac{12 u H(1,0,y) \
t}{s^2}+\frac{6 H(1,0,y) t}{{\tpu}}-\frac{4 u H(1,0,z) t}{s \
{\tpu}}+\frac{12 u H(1,0,z) t}{s^2}+\frac{4 u H(2,y) H(1,0,z) \
t}{{\tpu}^2}-\frac{2 H(1,0,z) t}{{\tpu}}+\frac{6 H(3,2,y) \
t}{u}+\frac{20 u H(0,1,0,y) t}{{\tpu}^2}-\frac{4 u H(0,1,0,z) \
t}{{\tpu}^2}+\frac{4 H(0,1,0,z) t}{s}-\frac{8 H(1,1,0,y) t}{s}-\frac{8 u \
H(1,1,0,z) t}{{\tpu}^2}-\frac{20 u H(2,1,0,y) t}{{\tpu}^2}
\end{dmath*}
\begin{dmath*}
{\white=}
-18 \
H(1,z)+9 H(1,z) H(2,y)-18 H(2,y)+8 H(1,z) H(3,y)+12 H(3,y) H(0,1,z)+4 \
H(0,1,z)+12 H(1,z) H(0,2,y)-4 H(0,1,z) H(0,2,y)-4 H(0,2,y)-12 H(1,z) \
H(0,3,y)-\frac{2 u^2 H(1,0,y)}{s {\tpu}}+\frac{8 u \
H(1,0,y)}{{\tpu}}-6 H(1,z) H(1,0,y)
\end{dmath*}
\begin{dmath*}
{\white=}
+4 H(0,1,z) H(1,0,y)+\frac{2 u^2 \
H(1,0,z)}{s {\tpu}}+\frac{2 u H(1,0,z)}{{\tpu}}-12 H(3,y) H(1,0,z)-8 \
H(0,2,y) H(1,0,z)-8 H(0,3,y) H(1,0,z)-24 H(3,y) H(1,1,z)+8 H(0,0,y) \
H(1,1,z)-8 H(0,3,y) H(1,1,z)+9 H(1,1,z)-4 H(0,1,z) H(1,2,y)+12 H(1,z) \
H(2,0,y)-4 H(0,1,z) H(2,0,y)-4 H(2,0,y)+8 H(0,0,z) H(2,2,y)+16 H(0,1,z) \
H(2,2,y)+9 H(2,2,y)-12 H(1,z) H(2,3,y)-12 H(1,z) H(3,0,y)-8 H(1,1,z) \
H(3,0,y)-24 H(1,z) H(3,2,y)+8 H(0,1,z) H(3,2,y)+8 H(3,2,y)+24 H(1,z) \
H(3,3,y)-8 H(0,1,z) H(3,3,y)+8 H(1,0,z) H(3,3,y)+16 H(1,1,z) H(3,3,y)-4 \
H(1,y) H(0,0,1,z)+8 H(2,y) H(0,0,1,z)+8 H(1,z) H(0,0,2,y)-8 H(1,z) \
H(0,0,3,y)-\frac{2 u^2 H(0,1,0,y)}{s^2}+\frac{12 u^2 \
H(0,1,0,y)}{{\tpu}^2}+\frac{4 u H(0,1,0,y)}{s}-4 H(1,z) H(0,1,0,y)+4 \
H(2,y) H(0,1,0,z)+8 H(3,y) H(0,1,1,z)-12 H(0,1,1,z)+8 H(1,z) H(0,2,0,y)+12 \
H(0,2,2,y)-4 H(1,z) H(0,2,3,y)
\end{dmath*}
\begin{dmath*}
{\white=}
-8 H(1,z) H(0,3,2,y)-12 H(0,3,2,y)+8 H(1,z) \
H(0,3,3,y)-8 H(1,z) H(1,0,0,y)+2 H(0,y) \Big(\frac{H(1,0,z) \
u^2}{s^2}-\frac{2 t (s-3 {\tpu}) H(0,z) u}{s^2 {\tpu}}-\frac{2 \
H(1,0,z) u}{s}+\frac{2 t H(1,0,z) u}{{\tpu}^2}+\frac{6 \
u}{s}+\Big(-\frac{3 u}{t}-2\Big) H(1,z)+6 H(1,1,z)-4 H(0,1,1,z)-2 \
H(1,0,1,z)\Big)-4 H(1,y) H(1,0,1,z)+12 H(2,y) H(1,0,1,z)+8 H(3,y) \
H(1,0,1,z)-6 H(1,0,1,z)-6 H(1,0,2,y)+4 H(1,z) H(1,0,3,y)+\frac{4 u^2 \
H(1,1,0,z)}{s^2}-\frac{6 u^2 H(1,1,0,z)}{{\tpu}^2}-\frac{8 u \
H(1,1,0,z)}{s}+4 H(2,y) H(1,1,0,z)-6 H(1,2,0,y)-4 H(1,z) H(1,2,3,y)+8 H(1,z) \
H(2,0,0,y)+12 H(2,0,2,y)-\frac{12 u^2 H(2,1,0,y)}{{\tpu}^2}-4 H(1,z) \
H(2,1,0,y)+12 H(2,2,0,y)+16 H(1,z) H(2,2,3,y)-12 H(2,3,2,y)-8 H(1,z) \
H(3,0,2,y)-12 H(3,0,2,y)+8 H(1,z) H(3,0,3,y)-8 H(1,z) H(3,2,0,y)-12 \
H(3,2,0,y)-24 H(3,2,2,y)+16 H(1,z) H(3,2,3,y)+8 H(1,z) H(3,3,0,y)+16 H(1,z) \
H(3,3,2,y)+24 H(3,3,2,y)+2 H(0,z) \Big(\frac{3 H(2,0,y) \
t^2}{{\tpu}^2}+\frac{{\spt} H(1,0,y) t}{s^2}+\frac{8 u H(2,0,y) \
t}{{\tpu}^2}+\frac{6 t}{s}+\Big(-\frac{3 t}{u}-2\Big) H(2,y)+6 \
H(0,2,y)-\frac{3 {\spt} H(1,0,y)}{s}
\end{dmath*}
\begin{dmath*}
{\white=}
+\frac{3 u^2 \
H(2,0,y)}{{\tpu}^2}+6 H(2,2,y)-6 H(3,2,y)+4 H(0,0,2,y)-4 H(0,3,2,y)-2 \
H(2,0,2,y)-4 H(3,0,2,y)-4 H(3,2,2,y)+4 H(3,3,2,y)\Big)-16 H(1,z) \
H(3,3,3,y)+8 H(0,0,1,1,z)+8 H(0,0,2,2,y)-8 H(0,0,3,2,y)+8 H(0,1,0,1,z)-4 \
H(0,1,0,2,y)+8 H(0,1,1,0,y)+8 H(0,1,1,0,z)-4 H(0,1,2,0,y)+8 H(0,2,0,2,y)-4 \
H(0,2,1,0,y)+8 H(0,2,2,0,y)-4 H(0,2,3,2,y)-8 H(0,3,2,2,y)+8 H(0,3,3,2,y)+4 \
H(1,0,0,1,z)-8 H(1,0,0,2,y)+4 H(1,0,1,0,y)-8 H(1,0,2,0,y)+4 H(1,0,3,2,y)+8 \
H(1,1,0,0,y)+8 H(1,1,0,1,z)+8 H(1,1,1,0,y)+8 H(1,1,1,0,z)-8 H(1,2,0,0,y)-4 \
H(1,2,1,0,y)-4 H(1,2,3,2,y)+8 H(2,0,0,2,y)-4 H(2,0,1,0,y)+8 H(2,0,2,0,y)-8 \
H(2,1,0,0,y)-4 H(2,1,0,2,y)+8 H(2,1,1,0,y)-4 H(2,1,2,0,y)+8 H(2,2,0,0,y)+16 \
H(2,2,3,2,y)-8 H(3,0,1,0,y)-8 H(3,0,2,2,y)+8 H(3,0,3,2,y)-8 H(3,2,0,2,y)+8 \
H(3,2,1,0,y)-8 H(3,2,2,0,y)+16 H(3,2,3,2,y)+8 H(3,3,0,2,y)+8 H(3,3,2,0,y)+16 \
H(3,3,2,2,y)-16 H(3,3,3,2,y)+6+\frac{6 u H(1,z) H(3,y)}{t}+\frac{6 u \
H(0,1,z)}{t}-\frac{6 u H(0,2,y)}{t}+\frac{6 u^2 H(1,0,y)}{{\tpu} \
t}-\frac{6 u H(2,0,y)}{t}+\frac{6 u H(3,2,y)}{t}
\Bigg)
\Bigg\}
\end{dmath*}
\begin{dmath*}
{\white=}
+
C_A n_f  \Bigg\{
\zeta_3 \Big( -\frac{37}{18} \Big)
+
\zeta_2 \Big(
-\frac{1}{3} H(0,y)+\frac{1}{6} H(2,y)-\frac{1}{3} H(0,z)+\frac{1}{6} \
H(1,z)+\frac{t u}{3 s^2}-\frac{7}{36}
\Big)
+
\Bigg(
\frac{1}{36} H(0,y) \Big(H(0,z) \Big(\frac{12 t u}{s^2}+20\Big)+31 \
H(1,z)-36 H(0,0,z)+24 H(0,1,z)-36 H(1,0,z)+24 H(1,1,z)+\frac{12 \
u}{s}+86\Big)+\frac{t u H(1,0,y)}{3 s^2}+\frac{t u H(1,0,z)}{3 s^2}+H(0,z) \
\Big(\frac{31}{36} H(2,y)-H(0,0,y)-\frac{1}{3} \
H(0,2,y)-H(2,0,y)+\frac{2}{3} H(2,2,y)+H(3,2,y)+\frac{t}{3 \
s}+\frac{43}{18}\Big)+H(2,y) H(1,z)-\frac{20}{9} H(3,y) H(1,z)-H(0,0,y) \
H(1,z)+\frac{2}{3} H(0,2,y) H(1,z)+H(0,3,y) H(1,z)+\frac{2}{3} H(2,0,y) \
H(1,z)-\frac{4}{3} H(2,3,y) H(1,z)+H(3,0,y) H(1,z)-\frac{4}{3} H(3,2,y) \
H(1,z)-\frac{4}{3} H(3,3,y) H(1,z)-H(2,y) H(0,0,z)-\frac{2}{3} H(2,y) \
H(0,1,z)-\frac{1}{3} H(3,y) H(0,1,z)+H(3,y) H(1,0,z)-\frac{4}{3} H(3,y) \
H(1,1,z)+\frac{17}{27} H(2,y)-\frac{41}{18} H(0,0,y)+\frac{31}{36} \
H(0,2,y)+\frac{31}{36} H(2,0,y)+H(2,2,y)-\frac{20}{9} \
H(3,2,y)-H(0,0,2,y)-H(0,2,0,y)+\frac{2}{3} \
H(0,2,2,y)+H(0,3,2,y)-H(2,0,0,y)+\frac{2}{3} H(2,0,2,y)-\frac{2}{3} \
H(2,1,0,y)+\frac{2}{3} H(2,2,0,y)-\frac{4}{3} \
H(2,3,2,y)+H(3,0,2,y)+H(3,2,0,y)-\frac{4}{3} H(3,2,2,y)-\frac{4}{3} \
H(3,3,2,y)+\frac{17}{27} H(1,z)-\frac{41}{18} H(0,0,z)-\frac{49}{36} \
H(0,1,z)+\frac{31}{36} H(1,0,z)+H(1,1,z)-\frac{1}{3} H(0,0,1,z)-\frac{2}{3} \
H(0,1,1,z)-H(1,0,0,z)-\frac{2}{3} H(1,0,1,z)+\frac{65}{162}
\Bigg)
\Bigg\}
\end{dmath*}
% \end{dgroup*}
% % 
% \begin{dgroup*}
\begin{dmath*}
{\white=}
+
C_F n_f  \Bigg\{
\zeta_3 \Big( -\frac{1}{9} \Big)
+
\zeta_2 \Big(   
\frac{1}{3} (4 H(1,y)-5 H(2,y)-H(1,z)+1)
\Big)
+
\Bigg(
-\frac{279}{2} H(0,y) H(1,z)-108 H(0,y) H(0,1,z)+27 H(0,y) H(1,0,z)-108 \
H(0,y) H(1,1,z)-\frac{279}{2} H(2,y) H(0,z)-162 H(2,y) H(1,z)+360 H(3,y) \
H(1,z)+162 H(0,0,y) H(1,z)+162 H(2,y) H(0,0,z)+108 H(2,y) H(0,1,z)+54 H(3,y) \
H(0,1,z)+54 H(0,2,y) H(0,z)-108 H(0,2,y) H(1,z)-162 H(0,3,y) H(1,z)-27 \
H(1,0,y) H(0,z)-108 H(2,y) H(1,0,z)-162 H(3,y) H(1,0,z)+216 H(3,y) \
H(1,1,z)+54 H(2,0,y) H(0,z)-108 H(2,0,y) H(1,z)-108 H(2,2,y) H(0,z)+216 \
H(2,3,y) H(1,z)-162 H(3,0,y) H(1,z)-162 H(3,2,y) H(0,z)+216 H(3,2,y) \
H(1,z)+216 H(3,3,y) H(1,z)-27 H(0,y)-102 H(2,y)-\frac{279}{2} H(0,2,y)+180 \
H(1,0,y)-\frac{279}{2} H(2,0,y)-162 H(2,2,y)+360 H(3,2,y)+162 H(0,0,2,y)-27 \
H(0,1,0,y)+162 H(0,2,0,y)-108 H(0,2,2,y)-162 H(0,3,2,y)-162 H(1,0,0,y)+108 \
H(1,1,0,y)+162 H(2,0,0,y)-108 H(2,0,2,y)-108 H(2,2,0,y)+216 H(2,3,2,y)-162 \
H(3,0,2,y)-162 H(3,2,0,y)+216 H(3,2,2,y)+216 H(3,3,2,y)-27 H(0,z)-102 \
H(1,z)+\frac{441}{2} H(0,1,z)+\frac{81}{2} H(1,0,z)-162 H(1,1,z)+54 \
H(0,0,1,z)-27 H(0,1,0,z)+108 H(0,1,1,z)+108 H(1,0,1,z)+200
\Bigg) \Big/ 81
\Bigg\}
\end{dmath*}
\begin{dmath*}
{\white=}
+
n_f^2  \Bigg\{
\frac{1}{108} \Big( H(0,y) (3 H(0,z)-20)+15 H(0,0,y)-20 H(0,z)+15 H(0,0,z)+6 \
\zeta_2 \Big)
\Bigg\}
\end{dmath*}
\end{dgroup*}
%
%%%%%% C2 term %%%%%
\begin{dgroup*}
\begin{dmath*}
 {\cal B}^{(2)}_{2 ; 2} = \frac{1}{24} \Big(121 {C_A}^2+44 {C_A} (6 {C_F}-{n_f})+4 \
\Big(27 {C_F}^2-12 {C_F} {n_f}+{n_f}^2\Big)\Big)  
\end{dmath*}
\begin{dmath*}
{\cal B}^{(2)}_{2 ; 1} = 
C_A^2 \Bigg\{
\frac{1}{108} \Big( 594 H(0,y) H(0,z)+594 H(0,y) H(1,z)+594 H(2,y) H(0,z)-1188 \
H(3,y) H(1,z)+990 H(0,y)+891 H(2,y)+594 H(0,2,y)+594 H(2,0,y)-1188 \
H(3,2,y)+990 H(0,z)+891 H(1,z)-594 H(0,1,z)+594 H(1,0,z)+495 \
\zeta_2-108 \zeta_3 - 1296 \Big)
\Bigg\}
+
C_A C_F \Bigg\{
\frac{1}{108} \Big(-1188 H(0,y) H(1,z)+54 H(0,y) (6 H(0,z)+6 H(1,z)+10)-1188 \
H(2,y) H(0,z)+54 (6 H(2,y)+10) H(0,z)+1728 H(3,y) H(1,z)-1296 H(2,y)-864 \
H(0,2,y)+1188 H(1,0,y)-864 H(2,0,y)+1728 H(3,2,y)-1296 H(1,z)+864 \
H(0,1,z)+324 H(1,0,z)+1566 {\zeta_2}-2808 {\zeta_3}-778  \Big)
\Bigg\}
+
C_F^2  \Bigg\{
\frac{1}{108} \Big( -648 H(0,y) H(1,z)-648 H(2,y) H(0,z)+1296 H(3,y) H(1,z)-972 \
H(2,y)-648 H(0,2,y)+648 H(1,0,y)-648 H(2,0,y)+1296 H(3,2,y)-972 H(1,z)+648 \
H(0,1,z)
\end{dmath*}
\begin{dmath*}
{\white=}
-1296 {\zeta_2}+2592 {\zeta_3}+972 \Big)
\Bigg\}
+
C_A n_f \Bigg\{
\frac{1}{108} \Big( -2 \Big( -27 (4 H(3,y)-3) H(1,z)+81 H(2,y)+54 H(0,2,y)+54 \
H(2,0,y)-108 H(3,2,y)-54 H(0,1,z)+54 H(1,0,z)+45 {\zeta_2}-260 \Big)-9 H(0,y) \
(12 H(0,z)+12 H(1,z)+31)-9 (12 H(2,y)+31) H(0,z) \Big)
\Bigg\}
+
C_F n_f \Bigg\{
\frac{1}{108} \Big( -2 \Big(54 ( 4 H(3,y)-3 ) H(1,z)-162 H(2,y)-108 H(0,2,y)+108 \
H(1,0,y)-108 H(2,0,y)+216 H(3,2,y)+108 H(0,1,z)+54 {\zeta_2}+22 \Big)-9 H(0,y) \
(6-24 H(1,z))-9 (6-24 H(2,y)) H(0,z) \Big)
\Bigg\}
+
n_f^2  \Bigg\{
\frac{1}{6} H(0,y)+\frac{1}{6} H(0,z)-\frac{10}{27}
\Bigg\}
\end{dmath*}
\begin{dmath*}
{\cal B}^{(2)}_{2 ; 0} = 
C_A^2  \Bigg\{
\zeta_4 \Bigg( \frac{39}{8} \Bigg)
+
\zeta_3 \Bigg(
-H(0,y)-6 H(1,y)+11 H(2,y)-H(0,z)+5 H(1,z)- \Big( 25 s^4 {\tpu}+2 s^3 \
\Big(79 t^2+266 t u+79 u^2\Big)+s^2 \Big(133 t^3+1097 t^2 u+1097 t \
u^2+133 u^3\Big)+2 s t u \Big(241 t^2+806 t u+241 u^2\Big)+457 t^2 \
{\tpu} u^2 \Big) \Big/ \Big( 36 {\spt}^2 {\spu}^2 {\tpu} \Big)
\Bigg)
+
\zeta_2 \Bigg(
\Big( H(0,z) \Big( s^2 (109 t+49 u)+2 s t (115 t+61 u)+t^2 (121 t+85 \
u)\Big)\Big) \Big/ \Big( 12 {\spt}^2 {\tpu} \Big)
+ \Big( H(0,y) \Big(s^2 (49 t+109 u)+2 \
s u (61 t+115 u)+u^2 (85 t+121 u)\Big) \Big) \Big/ \Big( 12 {\spu}^2 {\tpu} \Big)
+\Big( H(2,y) \Big(-144 s^5-257 s^4 {\tpu}-2 s^3 \Big(35 \
t^2+202 t u+35 u^2\Big)+s^2 \Big(43 t^3-25 t^2 u-25 t u^2+43 u^3\Big)+2 \
s t u \Big(55 t^2+74 t u+55 u^2\Big)+79 t^2 {\tpu} u^2\Big) \Big) \Big/ \Big( 12 {\spt}^2 {\spu}^2 {\tpu} \Big)
+\Big( H(1,z) \Big(-72 s^5-s^4 (113 \
t+41 u)+2 s^3 \Big(t^2+14 t u+73 u^2\Big)+s^2 \Big(43 t^3+191 t^2 u+407 \
t u^2+115 u^3\Big)+2 s t u \Big(55 t^2+182 t u+127 u^2\Big)+t^2 u^2 (79 \
t+151 u)\Big) \Big) \Big/ \Big( 12 {\spt}^2 {\spu}^2 {\tpu} \Big)
+\frac{6 {\spt} \
H(1,y)}{{\tpu}}+3 H(0,y) H(0,z)+2 H(1,y) H(0,z)+3 H(0,y) H(1,z)+2 H(1,y) \
H(1,z)+H(2,y) H(0,z)-2 H(3,y) H(1,z)+3 H(0,2,y)+2 H(1,2,y)+H(2,0,y)-2 \
H(3,2,y)+H(0,1,z)+H(1,0,z)+2 H(1,1,z)
-\Big( 28 s^3 {\tpu}+s^2 \Big(28 \
t^2+71 t u+28 u^2\Big)+34 s t {\tpu} u-3 t^2 u^2 \Big) \Big/ \Big( 9 s {\spt} {\spu} {\tpu} \Big)
\Bigg)
\end{dmath*}
\begin{dmath*}
{\white=}
+
\Bigg(
-\frac{2 (143 t+89 u) H(0,y)}{27 {\tpu}}
+\frac{1}{3} \Big(\frac{t u}{s \
{\tpu}}-17\Big) H(0,z) H(0,y)
-\Big( \Big(179 u^2+161 s u+107 t u+107 \
s t\Big) H(1,z) H(0,y) \Big) \Big/ \Big( 18 {\spu} {\tpu} \Big)
+7 H(0,0,z) H(0,y)
-\Big( \Big((13 t+25 u) s^2+2 u (19 t+28 u) s+31 {\tpu} \
u^2\Big) H(0,1,z) H(0,y)\Big) \Big/ \Big( 6 {\spu}^2 {\tpu} \Big)
+\Big( (17 t+23 u) \
H(1,0,z) H(0,y) \Big) \Big/ \Big( 2 {\tpu} \Big)
-\frac{2}{3} H(1,1,z) H(0,y)-2 H(0,0,1,z) \
H(0,y)-2 H(0,1,1,z) H(0,y)+2 H(1,0,0,z) H(0,y)+2 H(1,1,0,z) H(0,y)
-\frac{2 (89 t+143 u) H(0,z)}{27 {\tpu}}
-\frac{140}{27} H(1,z)
-\Big( \Big(179 t^2+161 s t+107 u t+107 s u\Big) H(0,z) H(2,y) \Big) \Big/ \Big( 18 {\spt} \
{\tpu} \Big) 
-\frac{13}{4} H(1,z) H(2,y)
-\frac{140}{27} H(2,y)
+\Big( \Big(188 \
s^2+197 {\tpu} s+206 t u\Big) H(1,z) H(3,y) \Big) \Big/ \Big(9 {\spt} \
{\spu} \Big) + 7 H(0,z) H(0,0,y)+7 H(1,z) H(0,0,y)
\end{dmath*}
\begin{dmath*}
{\white=}
+\frac{80}{9} H(0,0,y)+7 \
H(2,y) H(0,0,z)+2 H(0,0,y) H(0,0,z)+\frac{80}{9} H(0,0,z)
+\Big( \Big((215 \
t+269 u) s^2+\Big(215 t^2+520 u t+287 u^2\Big) s+t u (233 t+305 u)\Big) \
H(0,1,z)\Big) \Big/ \Big( 18 {\spt} {\spu} {\tpu}\Big) 
+\Big( \Big(18 s^5+59 \
{\tpu} s^4+\Big(61 t^2+152 u t+61 u^2\Big) s^3+4 \Big(5 t^3+31 u \
t^2+31 u^2 t+5 u^3\Big) s^2+34 t {\tpu}^2 u s+11 t^2 {\tpu} \
u^2\Big) H(2,y) H(0,1,z) \Big) \Big/ \Big( 3 {\spt}^2 {\spu}^2 \
{\tpu} \Big)
+\frac{1}{3} H(3,y) H(0,1,z)+2 H(0,0,y) H(0,1,z)
-\Big( \Big(179 \
u^2+161 s u+107 t u+107 s t\Big) H(0,2,y)\Big) \Big/ \Big( 18 {\spu} \
{\tpu} \Big) 
+ \Big( \Big((17 t+29 u) s^2+2 t (14 t+23 u) s+11 t^2 \
{\tpu}\Big) H(0,z) H(0,2,y) \Big) \Big/ \Big( 6 {\spt}^2 {\tpu} \Big) -\frac{2}{3} \
H(1,z) H(0,2,y)+2 H(0,0,z) H(0,2,y)-2 H(0,1,z) H(0,2,y)
-\Big( \Big((5 t+9 \
u) s^4+\Big(9 t^2+28 u t+19 u^2\Big) s^3+2 \Big(2 t^3+14 u t^2+21 u^2 \
t+5 u^3\Big) s^2+2 t u \Big(5 t^2+14 u t+9 u^2\Big) s+7 t^2 {\tpu} \
u^2\Big) H(1,z) H(0,3,y) \Big) \Big/ \Big( {\spt}^2 {\spu}^2 {\tpu} \Big)
-2 H(0,1,z) \
H(0,3,y)+\Big( u \Big(6 s^2-2 t s+9 u s+t u\Big) H(1,0,y)\Big) \Big/ \Big( 3 s {\spu} \
{\tpu} \Big) - \frac{3 u H(0,z) H(1,0,y)}{{\tpu}} 
- \Big( 3 (s+t+u) \
H(1,z) H(1,0,y) \Big) \Big/ \Big( {\tpu} \Big)
-\Big( \Big((125 t+107 u) s^2+t (125 t+119 u) \
s-6 t^2 u\Big) H(1,0,z) \Big) \Big/ \Big( 18 s {\spt} {\tpu} \Big)
+\Big( \Big(-3 \
s^3-(t-3 u) s^2+8 t {\tpu} s+6 t^2 {\tpu}\Big) H(2,y) \
H(1,0,z) \Big) \Big/ \Big( {\spt}^2 {\tpu} \Big) -7 H(3,y) H(1,0,z)+2 H(0,0,y) H(1,0,z)-6 \
H(0,3,y) H(1,0,z)
\end{dmath*}
\end{dgroup*}
%
%%%%
\begin{dgroup*}
\begin{dmath*}
{\white=}
+\frac{4}{3} H(3,y) H(1,1,z)+2 H(0,0,y) \
H(1,1,z)-\frac{13}{4} H(1,1,z)-2 H(0,1,z) H(1,2,y)+2 H(1,0,z) \
H(1,2,y)
-\Big( \Big(179 u^2+161 s u+107 t u+107 s t\Big) H(2,0,y) \Big) \Big/ \Big( 18 \
{\spu} {\tpu} \Big) + \frac{23}{2} H(0,z) H(2,0,y)-\frac{2}{3} H(1,z) \
H(2,0,y)+2 H(0,0,z) H(2,0,y)+2 H(0,1,z) H(2,0,y)+2 H(1,0,z) \
H(2,0,y)-\frac{2}{3} H(0,z) H(2,2,y)+2 H(0,0,z) H(2,2,y)+4 H(0,1,z) \
H(2,2,y)+4 H(1,0,z) H(2,2,y)-\frac{13}{4} H(2,2,y)
+\Big( \Big(18 s^5+61 \
{\tpu} s^4+5 \Big(13 t^2+32 u t+13 u^2\Big) s^3+2 \Big(11 t^3+67 u \
t^2+67 u^2 t+11 u^3\Big) s^2+38 t {\tpu}^2 u s+13 t^2 {\tpu} \
u^2\Big) H(1,z) H(2,3,y) \Big) \Big/ \Big( 3 {\spt}^2 {\spu}^2 {\tpu} \Big)-6 \
H(0,1,z) H(2,3,y)+2 H(1,0,z) H(2,3,y)-7 H(1,z) H(3,0,y)+2 H(0,1,z) \
H(3,0,y)-2 H(1,0,z) H(3,0,y)+ \Big( \Big(188 s^2+197 {\tpu} s+206 t \
u\Big) H(3,2,y) \Big) \Big/ \Big( 9 {\spt} {\spu} \Big) -7 H(0,z) H(3,2,y)+\frac{4}{3} \
H(1,z) H(3,2,y)+\frac{22}{3} H(1,z) H(3,3,y)-12 H(0,1,z) H(3,3,y)+4 H(1,0,z) \
H(3,3,y)+ \Big( \Big((7 u-5 t) s^4+\Big(-13 t^2+4 u t+17 u^2\Big) s^3-2 \
\Big(4 t^3+8 u t^2-13 u^2 t-5 u^3\Big) s^2+2 t u \Big(-5 t^2+2 u t+7 \
u^2\Big) s+t^2 {\tpu} u^2\Big) H(0,0,1,z) \Big) \Big/ \Big( 3 {\spt}^2 \
{\spu}^2 {\tpu} \Big) - 2 H(1,y) H(0,0,1,z)-2 H(2,y) H(0,0,1,z)-8 H(3,y) \
H(0,0,1,z)+2 H(0,z) H(0,0,2,y)+2 H(1,z) H(0,0,2,y)+7 H(0,0,2,y)
- \Big( u \Big(s^2+(u-2 t) s-3 t u\Big) H(0,1,0,y) \Big) \Big/ \Big( {\spu}^2 \
{\tpu} \Big) - \Big( t \Big(s^2+(t-2 u) s-3 t u\Big) \
H(0,1,0,z)\Big) \Big/ \Big( {\spt}^2 {\tpu} \Big) +2 H(1,y) H(0,1,0,z)+6 H(2,y) \
H(0,1,0,z)+4 H(3,y) H(0,1,0,z)+\frac{2}{3} H(0,1,1,z)+2 H(0,z) H(0,2,0,y)+2 \
H(1,z) H(0,2,0,y)+7 H(0,2,0,y)-2 H(0,z) H(0,2,2,y)-\frac{2}{3} H(0,2,2,y)+4 \
H(1,z) H(0,3,0,y)- \Big( \Big((5 t+9 u) s^4+\Big(9 t^2+28 u t+19 u^2\Big) \
s^3+2 \Big(2 t^3+14 u t^2+21 u^2 t+5 u^3\Big) s^2+2 t u \Big(5 t^2+14 u \
t+9 u^2\Big) s+7 t^2 {\tpu} u^2\Big) H(0,3,2,y) \Big) \Big/ \Big( {\spt}^2 \
{\spu}^2 {\tpu} \Big) - 6 H(0,z) H(0,3,2,y)+4 H(1,z) H(0,3,3,y)+2 H(2,y) \
H(1,0,0,z)
\end{dmath*}
\begin{dmath*}
{\white=}
+7 H(1,0,0,z)
+\Big( \Big(9 s^5+32 {\tpu} s^4+\Big(34 t^2+80 \
u t+34 u^2\Big) s^3+\Big(11 t^3+61 u t^2+61 u^2 t+11 u^3\Big) s^2+t u \
\Big(16 t^2+23 u t+16 u^2\Big) s+2 t^2 {\tpu} u^2\Big) \
H(1,0,1,z) \Big) \Big/ \Big( 3 {\spt}^2 {\spu}^2 {\tpu} \Big) -2 H(1,y) H(1,0,1,z)+6 \
H(2,y) H(1,0,1,z)
- \Big( 3 (s+t+u) H(1,0,2,y) \Big) \Big/ \Big({\tpu} \Big) -2 H(0,z) \
H(1,0,2,y)+2 H(1,z) H(1,0,3,y)+ \Big( 6 {\spt} \
H(1,1,0,y) \Big) \Big/ \Big( {\tpu} \Big) + \Big( \Big(2 (t+3 u) s^2+t (5 t+14 u) s+3 t^2 (t+3 \
u)\Big) H(1,1,0,z) \Big) \Big/ \Big( {\spt}^2 {\tpu} \Big) +2 H(1,y) H(1,1,0,z)+6 H(2,y) \
H(1,1,0,z)-\frac{3 (s+t+u) H(1,2,0,y)}{{\tpu}}-2 H(1,z) \
H(1,2,3,y)+2 H(0,z) H(2,0,0,y)+2 H(1,z) H(2,0,0,y)+7 H(2,0,0,y)-\frac{2}{3} \
H(2,0,2,y)+2 H(1,z) H(2,0,3,y)+ \Big( \Big(-9 s^3+(11 t-u) s^2+28 {\tpu} \
u s+20 {\tpu} u^2\Big) H(2,1,0,y) \Big) \Big/ \Big( 3 {\spu}^2 {\tpu} \Big) +2 H(1,z) \
H(2,1,0,y)+2 H(0,z) H(2,2,0,y)-\frac{2}{3} H(2,2,0,y)+4 H(1,z) H(2,2,3,y)+2 \
H(1,z) H(2,3,0,y)+ \Big( \Big(18 s^5+61 {\tpu} s^4+5 \Big(13 t^2+32 u \
t+13 u^2\Big) s^3+2 \Big(11 t^3+67 u t^2+67 u^2 t+11 u^3\Big) s^2+38 t \
{\tpu}^2 u s+13 t^2 {\tpu} u^2\Big) H(2,3,2,y) \Big) \Big/ \Big( 3 {\spt}^2 \
{\spu}^2 {\tpu} \Big) +2 H(0,z) H(2,3,2,y)-8 H(1,z) H(2,3,3,y)-2 H(0,z) \
H(3,0,2,y)-7 H(3,0,2,y)+4 H(1,z) H(3,0,3,y)-2 H(0,z) H(3,2,0,y)-7 \
H(3,2,0,y)+\frac{4}{3} H(3,2,2,y)+4 H(1,z) H(3,3,0,y)+4 H(0,z) \
H(3,3,2,y)+\frac{22}{3} H(3,3,2,y)-16 H(1,z) H(3,3,3,y)+8 H(0,0,1,0,z)+2 \
H(0,0,1,1,z)+2 H(0,0,2,2,y)+2 H(0,1,0,1,z)+6 H(0,1,1,0,z)+2 H(0,2,0,2,y)+4 \
H(0,2,1,0,y)+2 H(0,2,2,0,y)+4 H(0,3,0,2,y)+4 H(0,3,2,0,y)+4 H(0,3,3,2,y)+6 \
H(1,0,1,0,z)+2 H(1,0,3,2,y)+2 H(1,1,0,0,z)+4 H(1,1,0,1,z)+6 H(1,1,1,0,z)-2 \
H(1,2,3,2,y)+2 H(2,0,0,2,y)+2 H(2,0,1,0,y)+2 H(2,0,2,0,y)+2 H(2,0,3,2,y)+2 \
H(2,1,0,2,y)+2 H(2,1,2,0,y)+2 H(2,2,0,0,y)+4 H(2,2,1,0,y)+4 H(2,2,3,2,y)+2 \
H(2,3,0,2,y)+2 H(2,3,2,0,y)-8 H(2,3,3,2,y)+4 H(3,0,3,2,y)+4 H(3,3,0,2,y)+4 \
H(3,3,2,0,y)-16 H(3,3,3,2,y)+\frac{761}{81}
\Bigg)
\Bigg\}
\end{dmath*}
%%%% Ca Cf  
\begin{dmath*}
{\white=}
+
C_A C_F  \Bigg\{
\zeta_4 \Bigg( \frac{93}{4} \Bigg)
+
\zeta_3 \Bigg(
22 H(1,y)-18 H(2,y)+4 H(1,z)+ \Big( 803 s^5 {\tpu}+4 s^4 \Big(424 \
t^2+965 t u+424 u^2\Big)+s^3 \Big(821 t^3+5329 t^2 u+5329 t u^2+821 \
u^3\Big)+s^2 \Big(-72 t^4+1966 t^3 u+5336 t^2 u^2+1966 t u^3-72 \
u^4\Big)+s t u \Big(-144 t^3+1307 t^2 u+1307 t u^2-144 u^3\Big)-72 t^2 \
u^2 \Big(t^2+u^2\Big) \Big) \Big/ \Big( 18 s {\spt}^2 {\spu}^2 {\tpu} \Big)
\Bigg)
+
\zeta_2 \Bigg(
-\Big( 2 H(1,y) \Big(18 s^2+23 s t+11 s u-6 t^2\Big)  \Big) \Big/ \Big(  3 s \
{\tpu} \Big)
+\Big( t H(0,z) \Big(-10 s^3-3 s^2 (7 t+2 u)-9 s t {\tpu}+2 \
t^3\Big)  \Big) \Big/ \Big(  s {\spt}^2 {\tpu} \Big)
+\Big( u H(0,y) \Big(-10 s^3-3 s^2 (2 \
t+7 u)-9 s {\tpu} u+2 u^3\Big)  \Big) \Big/ \Big(  s {\spu}^2 \
{\tpu} \Big)
+\Big( H(2,y) \Big(144 s^5 {\tpu}+s^4 \Big(289 t^2+554 t \
u+289 u^2\Big)+4 s^3 \Big(32 t^3+147 t^2 u+147 t u^2+32 u^3\Big)+s^2 \
\Big(-17 t^4+90 t^3 u+166 t^2 u^2+90 t u^3-17 u^4\Big)-2 s t u \Big(35 \
t^3+111 t^2 u+111 t u^2+35 u^3\Big)-t^2 u^2 \Big(71 t^2+166 t u+71 \
u^2\Big)\Big)  \Big) \Big/ \Big(  6 {\spt}^2 {\spu}^2 {\tpu}^2 \Big)
-\Big( H(1,z) \
\Big(-72 s^6 {\tpu}^2-s^5 \Big(101 t^3+231 t^2 u+183 t u^2+53 \
u^3\Big)+4 s^4 {\tpu}^2 \Big(8 t^2+25 t u+26 u^2\Big)+s^3 \Big(61 \
t^5+427 t^4 u+1120 t^3 u^2+1216 t^2 u^3+523 t u^4+61 u^5\Big)+2 s^2 \
{\tpu}^2 u \Big(79 t^3+236 t^2 u+79 t u^2-12 u^3\Big)+s t u^2 \
\Big(115 t^4+369 t^3 u+321 t^2 u^2+19 t u^3-48 u^4\Big)-24 t^2 \
{\tpu}^2 u^4\Big)  \Big) \Big/ \Big(  6 s {\spt}^2 {\spu}^2 {\tpu}^3 \Big)
-2 \
H(1,y) H(0,z)-6 H(0,y) H(1,z)-2 H(1,y) H(1,z)-4 H(2,y) H(0,z)+2 H(0,2,y)+6 \
H(1,0,y)-2 H(1,2,y)-4 H(2,0,y)-8 H(2,1,y)+8 H(2,2,y)+2 H(0,1,z)+2 H(1,0,z)-2 \
H(1,1,z)
+\Big( 19 s^3 {\tpu}+s^2 \Big(19 t^2+182 t u+19 u^2\Big)+145 \
s t {\tpu} u+108 t^2 u^2  \Big) \Big/ \Big(  6 s {\spt} {\spu} {\tpu} \Big)
\Bigg)
\end{dmath*}
%%%% Ca Cf  
\begin{dmath*}
{\white=}
+
\Bigg(
\Big( \Big(26 t^2+26 s t+33 u t+15 s u\Big) H(0,y) \Big) \Big/ \Big( 3 {\spt} \
{\tpu} \Big) + \Big(\Big( 18 t u \Big) \Big/ \Big( s {\tpu} \Big) +3\Big) H(0,z) \
H(0,y)+ \Big( \Big(242 u^2+215 s u+134 t u+134 s t\Big) H(1,z) H(0,y) \Big) \Big/ \Big( 9 \
{\spu} {\tpu} \Big) + \Big( \Big((13 t+31 u) s^2+u (44 t+71 u) s+40 \
{\tpu} u^2\Big) H(0,1,z) H(0,y) \Big) \Big/ \Big(  3 {\spu}^2 \
{\tpu} \Big) - \Big( \Big(19 s t^2+62 s u t+31 s u^2-6 {\tpu} u^2\Big) \
H(1,0,z) H(0,y) \Big) \Big/ \Big(  3 s {\tpu}^2 \Big) - \frac{14}{3} H(1,1,z) H(0,y)+4 H(0,0,1,z) \
H(0,y)+2 H(0,1,0,z) H(0,y)+8 H(0,1,1,z) H(0,y)+2 H(1,0,1,z) H(0,y)-4 \
H(1,1,0,z) H(0,y)+ \Big( \Big(26 u^2+26 s u+33 t u+15 s t\Big) H(0,z) \Big) \Big/ \Big( 3 \
{\spu} {\tpu} \Big) + \frac{803}{54} H(1,z) + \Big( \Big(242 t^2+215 s t+134 \
u t+134 s u\Big) H(0,z) H(2,y) \Big) \Big/ \Big(  9 {\spt} {\tpu} \Big) +2 H(1,z) \
H(2,y)+\frac{803}{54} H(2,y) - \Big( \Big(466 s^2+493 {\tpu} s+520 t \
u\Big) H(1,z) H(3,y) \Big) \Big/ \Big( 9 {\spt} {\spu} \Big) - 14 H(1,z) H(0,0,y)-14 \
H(2,y) H(0,0,z)- \Big( \Big((251 t+332 u) s^2+\Big(251 t^2+637 u t+359 \
u^2\Big) s+2 t u (139 t+193 u)\Big) H(0,1,z) \Big) \Big/ \Big( 9 {\spt} {\spu} \
{\tpu} \Big) - \Big( \Big(36 s^5+112 {\tpu} s^4+\Big(107 t^2+268 u t+107 \
u^2\Big) s^3+\Big(31 t^3+191 u t^2+191 u^2 t+31 u^3\Big) s^2+2 t u \
\Big(22 t^2+35 u t+22 u^2\Big) s+4 t^2 {\tpu} u^2\Big) H(2,y) \
H(0,1,z) \Big) \Big/ \Big( 3 {\spt}^2 {\spu}^2 {\tpu} \Big) -\frac{20}{3} H(3,y) \
H(0,1,z)-4 H(0,0,y) H(0,1,z) + \Big( \Big(242 u^2+215 s u+134 t u+134 s \
t\Big) H(0,2,y) \Big) \Big/ \Big( 9 {\spu} {\tpu} \Big) - \Big( \Big((29 t+47 u) s^2+t \
(49 t+76 u) s+20 t^2 {\tpu}\Big) H(0,z) H(0,2,y) \Big) \Big/ \Big( 3 {\spt}^2 \
{\tpu} \Big) - \frac{14}{3} H(1,z) H(0,2,y)-4 H(0,0,z) H(0,2,y)+6 H(0,1,z) \
H(0,2,y)+ \Big( \Big(2 (7 t+13 u) s^4+5 \Big(5 t^2+16 u t+11 u^2\Big) \
s^3+\Big(11 t^3+79 u t^2+121 u^2 t+29 u^3\Big) s^2+4 t u \Big(7 t^2+20 u \
t+13 u^2\Big) s+20 t^2 {\tpu} u^2\Big) H(1,z) H(0,3,y) \Big) \Big/ \Big(  {\spt}^2 \
{\spu}^2 {\tpu} \Big) 
\end{dmath*}
%%%% Ca Cf  
\begin{dmath*}
{\white=}
+ 4 H(0,1,z) H(0,3,y)+ \Big( \Big(-2 (76 t+103 u) \
s^2+(37 t-233 u) u s+162 t u^2\Big) H(1,0,y) \Big) \Big/ \Big( 9 s {\spu} \
{\tpu} \Big) + \Big( 2 \Big(3 t^2+8 s t+14 s u\Big) H(0,z) H(1,0,y) \Big) \Big/ \Big( 3 s \
{\tpu} \Big) + \Big(\frac{6 s}{{\tpu}} + 9\Big) H(1,z) H(1,0,y)-2 H(0,1,z) \
H(1,0,y)+ \Big( \Big((t-2 u) s^2+t (t+19 u) s+18 t^2 u\Big) H(1,0,z) \Big) \Big/ \Big( s \
{\spt} {\tpu} \Big) + \Big( \Big(18 {\tpu} s^3+\Big(31 t^2+32 u t+13 \
u^2\Big) s^2-t \Big(t^2+17 u t-8 u^2\Big) s-2 t^2 \Big(7 t^2+20 u t+7 \
u^2\Big)\Big) H(2,y) H(1,0,z) \Big) \Big/ \Big( 3 {\spt}^2 {\tpu}^2 \Big) +20 H(3,y) \
H(1,0,z)-4 H(0,0,y) H(1,0,z)+8 H(0,2,y) H(1,0,z)+16 H(0,3,y) H(1,0,z)+2 \
H(1,0,y) H(1,0,z)+\frac{28}{3} H(3,y) H(1,1,z)-8 H(0,0,y) H(1,1,z)+4 \
H(0,3,y) H(1,1,z)+2 H(1,1,z)+6 H(0,1,z) H(1,2,y)-2 H(1,0,z) \
H(1,2,y)+ \Big( \Big(242 u^2+215 s u+134 t u+134 s t\Big) H(2,0,y) \Big) \Big/ \Big( 9 \
{\spu} {\tpu} \Big) - \Big( \Big(47 t^3+153 u t^2+153 u^2 t+47 u^3\Big) \
H(0,z) H(2,0,y) \Big) \Big/ \Big( 3 {\tpu}^3 \Big) - \frac{14}{3} H(1,z) H(2,0,y)-4 H(0,0,z) \
H(2,0,y)-2 H(0,1,z) H(2,0,y)-6 H(1,0,z) H(2,0,y)-\frac{14}{3} H(0,z) \
H(2,2,y)-8 H(0,0,z) H(2,2,y)-16 H(0,1,z) H(2,2,y)-8 H(1,0,z) H(2,2,y)+2 \
H(2,2,y)-\Big( \Big(36 s^5+98 {\tpu} s^4+\Big(79 t^2+212 u t+79 \
u^2\Big) s^3+\Big(17 t^3+121 u t^2+121 u^2 t+17 u^3\Big) s^2+2 t u \
\Big(8 t^2+7 u t+8 u^2\Big) s-10 t^2 {\tpu} u^2\Big) H(1,z) \
H(2,3,y) \Big) \Big/ \Big( 3 {\spt}^2 {\spu}^2 {\tpu} \Big) +12 H(0,1,z) H(2,3,y)-4 \
H(1,0,z) H(2,3,y)+20 H(1,z) H(3,0,y)-4 H(0,1,z) H(3,0,y)+4 H(1,1,z) \
H(3,0,y)- \Big( \Big(466 s^2+493 {\tpu} s+520 t u\Big) H(3,2,y) \Big) \Big/ \Big( 9 \
{\spt} {\spu} \Big) +20 H(0,z) H(3,2,y)+ \frac{28}{3} H(1,z) H(3,2,y)-4 \
H(0,1,z) H(3,2,y)-4 H(1,0,z) H(3,2,y)-\frac{80}{3} H(1,z) H(3,3,y)+28 \
H(0,1,z) H(3,3,y)-12 H(1,0,z) H(3,3,y)-8 H(1,1,z) H(3,3,y)+ \Big( \Big(4 (4 \
t-5 u) s^4+\Big(41 t^2-8 u t-49 u^2\Big) s^3+\Big(25 t^3+53 u t^2-73 u^2 \
t-29 u^3\Big) s^2+8 t u \Big(4 t^2-u t-5 u^2\Big) s-2 t^2 {\tpu} \
u^2\Big) H(0,0,1,z) \Big) \Big/ \Big( 3 {\spt}^2 {\spu}^2 {\tpu} \Big) 
\end{dmath*}
%%%% Ca Cf  
\begin{dmath*}
{\white=}
+6 H(1,y) \
H(0,0,1,z)+16 H(3,y) H(0,0,1,z)-8 H(0,z) H(0,0,2,y)-8 H(1,z) H(0,0,2,y)-14 \
H(0,0,2,y)+4 H(1,z) H(0,0,3,y)+ \Big( \Big(-6 {\tpu} u^4-s \Big(17 \
t^2+22 u t+17 u^2\Big) u^2+s^2 \Big(2 t^2+19 u t-7 u^2\Big) u+2 s^3 \
\Big(5 t^2+13 u t+2 u^2\Big)\Big) H(0,1,0,y) \Big) \Big/ \Big( 3 s {\spu}^2 \
{\tpu}^2 \Big) +2 H(0,z) H(0,1,0,y)+2 H(1,z) H(0,1,0,y)+ \Big( \Big(-6 \
{\tpu} t^4+s \Big(t^2+14 u t+u^2\Big) t^2+s^2 \Big(29 t^2+91 u t+38 \
u^2\Big) t+s^3 \Big(22 t^2+62 u t+28 u^2\Big)\Big) H(0,1,0,z) \Big) \Big/ \Big( 3 s \
{\spt}^2 {\tpu}^2 \Big) -2 H(1,y) H(0,1,0,z)-12 H(2,y) H(0,1,0,z)-4 H(3,y) \
H(0,1,0,z)-4 H(3,y) H(0,1,1,z)+\frac{14}{3} H(0,1,1,z)-8 H(1,z) \
H(0,2,0,y)-14 H(0,2,0,y)+4 H(0,z) H(0,2,2,y)- \frac{14}{3} H(0,2,2,y)+2 \
H(1,z) H(0,2,3,y)-8 H(1,z) H(0,3,0,y)+ \Big( \Big(2 (7 t+13 u) s^4+5 \Big(5 \
t^2+16 u t+11 u^2\Big) s^3+\Big(11 t^3+79 u t^2+121 u^2 t+29 u^3\Big) \
s^2+4 t u \Big(7 t^2+20 u t+13 u^2\Big) s+20 t^2 {\tpu} u^2\Big) \
H(0,3,2,y) \Big) \Big/ \Big( {\spt}^2 {\spu}^2 {\tpu} \Big) +16 H(0,z) H(0,3,2,y)+4 \
H(1,z) H(0,3,2,y)-12 H(1,z) H(0,3,3,y)+4 H(0,z) H(1,0,0,y)+4 H(1,z) \
H(1,0,0,y)+14 H(1,0,0,y)-4 H(2,y) H(1,0,0,z)- \Big( \Big(18 s^5+49 \
{\tpu} s^4+\Big(35 t^2+88 u t+35 u^2\Big) s^3+4 \Big(t^3+5 u t^2+5 \
u^2 t+u^3\Big) s^2-2 t u \Big(5 t^2+28 u t+5 u^2\Big) s-23 t^2 \
{\tpu} u^2\Big) H(1,0,1,z) \Big) \Big/ \Big( 3 {\spt}^2 {\spu}^2 {\tpu} \Big) +6 \
H(1,y) H(1,0,1,z)-18 H(2,y) H(1,0,1,z)-4 H(3,y) H(1,0,1,z)
+\Big( \frac{6 s}{{\tpu}} + 9\Big) H(1,0,2,y)+4 H(0,z) H(1,0,2,y)-6 H(1,z) \
H(1,0,3,y)- \Big( 2 \Big(18 s^2+23 t s+11 u s-6 t^2\Big) H(1,1,0,y) \Big) \Big/ \Big( 3 s \
{\tpu} \Big) - \Big( \Big(2 u (3 t+u) s^3+\Big(3 t^3+21 u t^2+6 u^2 t-4 \
u^3\Big) s^2+t \Big(3 t^3+18 u t^2+3 u^2 t-8 u^3\Big) s-4 t^2 \
{\tpu} u^2\Big) H(1,1,0,z) \Big) \Big/ \Big( s {\spt}^2 {\tpu}^2 \Big) -2 H(1,y) \
H(1,1,0,z)-18 H(2,y) H(1,1,0,z)-4 H(3,y) H(1,1,0,z) 
+ \Big(\frac{6 s}{{\tpu}}+9\Big) H(1,2,0,y)+2 H(0,z) H(1,2,0,y)
\end{dmath*}
%%%% Ca Cf  
\begin{dmath*}
{\white=}
+6 H(1,z) H(1,2,3,y)-4 \
H(0,z) H(2,0,0,y)-8 H(1,z) H(2,0,0,y)-14 H(2,0,0,y)+2 H(0,z) \
H(2,0,2,y)-\frac{14}{3} H(2,0,2,y)-4 H(1,z) H(2,0,3,y)+ \Big( \Big(6 \
{\tpu} s^3+\Big(9 t^2+20 u t+15 u^2\Big) s^2+u \Big(12 t^2+13 u t+9 \
u^2\Big) s-4 t u^3\Big) H(2,1,0,y) \Big) \Big/ \Big( {\spu}^2 {\tpu}^2 \Big) -2 H(0,z) \
H(2,1,0,y)-2 H(1,z) H(2,1,0,y)-4 H(0,z) H(2,2,0,y)-\frac{14}{3} \
H(2,2,0,y)-16 H(1,z) H(2,2,3,y)-4 H(1,z) H(2,3,0,y)- \Big( \Big(36 s^5+98 \
{\tpu} s^4+\Big(79 t^2+212 u t+79 u^2\Big) s^3+\Big(17 t^3+121 u \
t^2+121 u^2 t+17 u^3\Big) s^2+2 t u \Big(8 t^2+7 u t+8 u^2\Big) s-10 \
t^2 {\tpu} u^2\Big) H(2,3,2,y) \Big) \Big/ \Big( 3 {\spt}^2 {\spu}^2 \
{\tpu} \Big) -4 H(0,z) H(2,3,2,y)+16 H(1,z) H(2,3,3,y)+8 H(0,z) H(3,0,2,y)+4 \
H(1,z) H(3,0,2,y)+20 H(3,0,2,y)-12 H(1,z) H(3,0,3,y)+4 H(1,z) H(3,2,0,y)+20 \
H(3,2,0,y)+4 H(0,z) H(3,2,2,y)+\frac{28}{3} H(3,2,2,y)-8 H(1,z) \
H(3,2,3,y)-12 H(1,z) H(3,3,0,y)-12 H(0,z) H(3,3,2,y)-8 H(1,z) \
H(3,3,2,y)-\frac{80}{3} H(3,3,2,y)+40 H(1,z) H(3,3,3,y)+4 H(0,0,1,0,y)-8 \
H(0,0,1,0,z)-8 H(0,0,1,1,z)-8 H(0,0,2,2,y)+4 H(0,0,3,2,y)-8 H(0,1,0,1,z)+2 \
H(0,1,0,2,y)-12 H(0,1,1,0,z)+2 H(0,1,2,0,y)-8 H(0,2,0,2,y)-2 H(0,2,1,0,y)-8 \
H(0,2,2,0,y)+2 H(0,2,3,2,y)-8 H(0,3,0,2,y)-8 H(0,3,2,0,y)+4 H(0,3,2,2,y)-12 \
H(0,3,3,2,y)-2 H(1,0,0,1,z)+4 H(1,0,0,2,y)+4 H(1,0,1,0,y)-4 H(1,0,1,0,z)+4 \
H(1,0,2,0,y)-6 H(1,0,3,2,y)-12 H(1,1,0,1,z)-14 H(1,1,1,0,z)+4 H(1,2,0,0,y)+4 \
H(1,2,1,0,y)+6 H(1,2,3,2,y)-8 H(2,0,0,2,y)-8 H(2,0,2,0,y)-4 H(2,0,3,2,y)+4 \
H(2,1,0,0,y)-2 H(2,1,0,2,y)-8 H(2,1,1,0,y)-2 H(2,1,2,0,y)-8 H(2,2,0,0,y)-8 \
H(2,2,1,0,y)-16 H(2,2,3,2,y)-4 H(2,3,0,2,y)-4 H(2,3,2,0,y)+16 H(2,3,3,2,y)+8 \
H(3,0,1,0,y)+4 H(3,0,2,2,y)-12 H(3,0,3,2,y)+4 H(3,2,0,2,y)-8 H(3,2,1,0,y)+4 \
H(3,2,2,0,y)-8 H(3,2,3,2,y)-12 H(3,3,0,2,y)-12 H(3,3,2,0,y)-8 \
H(3,3,2,2,y)+40 H(3,3,3,2,y)-\frac{4003}{162}
\Bigg)
\Bigg\}
\end{dmath*}
%%%% Cf^2  
\begin{dmath*}
{\white=}
+
C_F^2  \Bigg\{
\zeta_4 \Bigg(  -22 \Bigg)
+
\zeta_3 \Bigg(
-16 H(1,y)+8 H(2,y)-8 H(1,z)-  \Big(  2 \Big(25 s^5 {\tpu}+s^4 \Big(51 \
t^2+112 t u+51 u^2\Big)+4 s^3 \Big(6 t^3+37 t^2 u+37 t u^2+6 u^3\Big)-2 \
s^2 \Big(t^4-27 t^3 u-69 t^2 u^2-27 t u^3+u^4\Big)+s t u \Big(-4 t^3+33 \
t^2 u+33 t u^2-4 u^3\Big)-2 t^2 u^2 \Big(t^2+u^2\Big)\Big) \Big) \Big/ \Big( s \
{\spt}^2 {\spu}^2 {\tpu} \Big)
\Bigg)
+
\zeta_2 \Bigg(
\Big( 2 t H(0,z) \Big(4 s^3+2 s^2 (4 t+u)+3 s t {\tpu}-t^3\Big) \Big) \Big/ \Big( s \
{\spt}^2 {\tpu} \Big) + \Big( 2 u H(0,y) \Big(4 s^3+2 s^2 (t+4 u)+3 s \
{\tpu} u-u^3\Big) \Big) \Big/ \Big( s {\spu}^2 {\tpu} \Big) + \Big( 2 H(2,y) \Big(s^4 \
\Big(5 t^2+12 t u+5 u^2\Big)+s^3 \Big(11 t^3+39 t^2 u+39 t u^2+11 \
u^3\Big)+s^2 \Big(6 t^4+42 t^3 u+76 t^2 u^2+42 t u^3+6 u^4\Big)+2 s t u \
\Big(7 t^3+26 t^2 u+26 t u^2+7 u^3\Big)+t^2 u^2 \Big(9 t^2+20 t u+9 \
u^2\Big)\Big) \Big) \Big/ \Big( {\spt}^2 {\spu}^2 {\tpu}^2 \Big) + \Big( 2 H(1,z) \
\Big(s^5 \Big(5 t^2+16 t u+9 u^2\Big)+s^4 \Big(11 t^3+47 t^2 u+53 t \
u^2+17 u^3\Big)+s^3 \Big(6 t^4+46 t^3 u+92 t^2 u^2+54 t u^3+6 \
u^4\Big)+2 s^2 u \Big(7 t^4+29 t^3 u+29 t^2 u^2+6 t u^3-u^4\Big)+s t \
u^2 \Big(9 t^3+20 t^2 u+5 t u^2-4 u^3\Big)-2 t^2 {\tpu} u^4\Big) \Big) \Big/ \Big( s \
{\spt}^2 {\spu}^2 {\tpu}^2 \Big) + \Big( 4 t (2 s-t) H(1,y) \Big) \Big/ \Big( s \
{\tpu} \Big) -8 H(2,y) H(1,z)+8 H(0,1,y)-8 H(0,2,y)+8 H(1,1,y)+8 H(2,1,y)-16 \
H(2,2,y)+ \Big( 2 \Big(6 s^3 {\tpu}+s^2 \Big(6 t^2+4 t u+6 u^2\Big)-s \
t {\tpu} u-6 t^2 u^2\Big) \Big) \Big/ \Big( s {\spt} {\spu} {\tpu} \Big)
\Bigg)
\end{dmath*}
%%%% Cf^2  
\begin{dmath*}
{\white=}
+
\Bigg(
- \Big( 12 t u H(0,y) H(0,z) \Big) \Big/ \Big( s {\tpu} \Big) - \Big( 2 (3 s (2 t+u)+t (7 t+3 u)) \
H(2,y) H(0,z) \Big) \Big/ \Big( {\spt} {\tpu} \Big) + \Big( 2 \Big((4 t+6 u) s^2+t (7 t+10 \
u) s+3 t^2 {\tpu}\Big) H(0,2,y) H(0,z) \Big) \Big/ \Big( {\spt}^2 \
{\tpu} \Big) - \Big( 2 \Big(t^2+s (t+3 u)\Big) H(1,0,y) H(0,z) \Big) \Big/ \Big( s \
{\tpu} \Big) + \Big( \Big(6 t^3+22 u t^2+22 u^2 t+6 u^3\Big) H(2,0,y) \
H(0,z) \Big) \Big/ \Big( {\tpu}^3 \Big) +12 H(2,2,y) H(0,z)-12 H(3,2,y) H(0,z)+8 H(0,0,2,y) \
H(0,z)-8 H(0,3,2,y) H(0,z)-4 H(2,0,2,y) H(0,z)-8 H(3,0,2,y) H(0,z)-8 \
H(3,2,2,y) H(0,z)+8 H(3,3,2,y) H(0,z)- \Big( 2 (3 s (t+2 u)+u (3 t+7 u)) \
H(0,y) H(1,z) \Big) \Big/ \Big( {\spu} {\tpu} \Big) -9 H(1,z)+9 H(1,z) H(2,y)-9 \
H(2,y)+ \Big( \Big(20 s^2+22 {\tpu} s+24 t u\Big) H(1,z) \
H(3,y) \Big) \Big/ \Big( {\spt} {\spu} \Big) + \Big( 2 \Big((4 t+7 u) s^2+\Big(4 t^2+13 u \
t+8 u^2\Big) s+t u (5 t+9 u)\Big) H(0,1,z) \Big) \Big/ \Big( {\spt} {\spu} \
{\tpu} \Big) - \Big( 2 u ({\spt}+u) (2 s+3 u) H(0,y) H(0,1,z) \Big) \Big/ \Big( {\spu}^2 \
{\tpu} \Big) - \Big( 2 \Big(2 {\tpu} s^4+\Big(5 t^2+12 u t+5 u^2\Big) \
s^3+\Big(3 t^3+19 u t^2+19 u^2 t+3 u^3\Big) s^2+2 t u \Big(4 t^2+11 u \
t+4 u^2\Big) s+6 t^2 {\tpu} u^2\Big) H(2,y) H(0,1,z) \Big) \Big/ \Big( {\spt}^2 \
{\spu}^2 {\tpu} \Big) +12 H(3,y) H(0,1,z) - \Big( 2 (3 s (t+2 u)+u (3 t+7 \
u)) H(0,2,y) \Big) \Big/ \Big( {\spu} {\tpu} \Big) + 12 H(1,z) H(0,2,y)-4 H(0,1,z) \
H(0,2,y)- \Big( 2 \Big(4 (t+2 u) s^4+\Big(7 t^2+24 u t+17 u^2\Big) \
s^3+\Big(3 t^3+23 u t^2+37 u^2 t+9 u^3\Big) s^2+8 t u \Big(t^2+3 u t+2 \
u^2\Big) s+6 t^2 {\tpu} u^2\Big) H(1,z) H(0,3,y) \Big) \Big/ \Big( {\spt}^2 \
{\spu}^2 {\tpu} \Big) + \Big( 2 \Big((3 t+5 u) s^2+\Big(6 u^2-4 t \
u\Big) s-6 t u^2\Big) H(1,0,y) \Big) \Big/ \Big( s {\spu} {\tpu} \Big) -6 H(1,z) \
H(1,0,y)+4 H(0,1,z) H(1,0,y) 
\end{dmath*}
%%%% Cf^2  
\begin{dmath*}
{\white=}
- \Big( 2 t \Big(s^2+(t+7 u) s+6 t u\Big) \
H(1,0,z) \Big) \Big/ \Big( s {\spt} {\tpu} \Big) + \Big( 2 u (2 s (2 t+u)-{\tpu} u) \
H(0,y) H(1,0,z) \Big) \Big/ \Big( s {\tpu}^2 \Big) + \Big( 2 t \Big(2 (t+2 u) s^2+\Big(5 \
t^2+11 u t+2 u^2\Big) s+t \Big(3 t^2+8 u t+3 u^2\Big)\Big) H(2,y) \
H(1,0,z) \Big) \Big/ \Big( {\spt}^2 {\tpu}^2 \Big) -12 H(3,y) H(1,0,z)-8 H(0,2,y) \
H(1,0,z)-8 H(0,3,y) H(1,0,z)+12 H(0,y) H(1,1,z)-24 H(3,y) H(1,1,z)+8 \
H(0,0,y) H(1,1,z)-8 H(0,3,y) H(1,1,z)+9 H(1,1,z)-4 H(0,1,z) H(1,2,y)- \Big( 2 \
(3 s (t+2 u)+u (3 t+7 u)) H(2,0,y) \Big) \Big/ \Big( {\spu} {\tpu} \Big) +12 H(1,z) \
H(2,0,y)-4 H(0,1,z) H(2,0,y)+8 H(0,0,z) H(2,2,y)+16 H(0,1,z) H(2,2,y)+9 \
H(2,2,y)
\end{dmath*}
%%%% Cf^2  
\begin{dmath*}
{\white=}
- \Big( 2 \Big(8 {\tpu} s^4+\Big(17 t^2+36 u t+17 u^2\Big) \
s^3+\Big(9 t^3+49 u t^2+49 u^2 t+9 u^3\Big) s^2+2 t u \Big(10 t^2+23 u \
t+10 u^2\Big) s+12 t^2 {\tpu} u^2\Big) H(1,z) H(2,3,y) \Big) \Big/ \Big( {\spt}^2 \
{\spu}^2 {\tpu} \Big) -12 H(1,z) H(3,0,y)-8 H(1,1,z) \
H(3,0,y)+ \Big( \Big(20 s^2+22 {\tpu} s+24 t u\Big) \
H(3,2,y) \Big) \Big/ \Big( {\spt} {\spu} \Big) -24 H(1,z) H(3,2,y)+8 H(0,1,z) H(3,2,y)+24 \
H(1,z) H(3,3,y)-8 H(0,1,z) H(3,3,y)+8 H(1,0,z) H(3,3,y)+16 H(1,1,z) \
H(3,3,y)- \Big( 2 s (t-u) \Big(2 s^3+5 {\tpu} s^2+\Big(3 t^2+10 u t+3 \
u^2\Big) s+4 t {\tpu} u\Big) H(0,0,1,z) \Big) \Big/ \Big( {\spt}^2 {\spu}^2 \
{\tpu} \Big) - 4 H(1,y) H(0,0,1,z)+8 H(2,y) H(0,0,1,z)+8 H(1,z) H(0,0,2,y)-8 \
H(1,z) H(0,0,3,y)+ \Big( 2 \Big({\tpu} u^4+s \Big(9 t^2+16 u t+9 \
u^2\Big) u^2+2 s^2 \Big(7 t^2+12 u t+7 u^2\Big) u+2 s^3 \Big(3 t^2+5 u \
t+3 u^2\Big)\Big) H(0,1,0,y) \Big) \Big/ \Big( s {\spu}^2 {\tpu}^2 \Big) -4 H(1,z) \
H(0,1,0,y)+ \Big( 2 t \Big(-2 u s^3+2 \Big(t^2+u^2\Big) s^2+t \Big(3 \
t^2+4 u t+3 u^2\Big) s+t^3 {\tpu}\Big) H(0,1,0,z) \Big) \Big/ \Big( s {\spt}^2 \
{\tpu}^2 \Big) +4 H(2,y) H(0,1,0,z)-8 H(0,y) H(0,1,1,z)+8 H(3,y) H(0,1,1,z)-12 \
H(0,1,1,z)+8 H(1,z) H(0,2,0,y)+12 H(0,2,2,y)-4 H(1,z) H(0,2,3,y)- \Big( 2 \
\Big(4 (t+2 u) s^4+\Big(7 t^2+24 u t+17 u^2\Big) s^3+\Big(3 t^3+23 u \
t^2+37 u^2 t+9 u^3\Big) s^2+8 t u \Big(t^2+3 u t+2 u^2\Big) s+6 t^2 \
{\tpu} u^2\Big) H(0,3,2,y) \Big) \Big/ \Big( {\spt}^2 {\spu}^2 {\tpu} \Big) -8 \
H(1,z) H(0,3,2,y)+8 H(1,z) H(0,3,3,y)-8 H(1,z) H(1,0,0,y)- \Big( 2 \Big(5 \
{\tpu} s^4+\Big(11 t^2+24 u t+11 u^2\Big) s^3+\Big(6 t^3+34 u t^2+34 \
u^2 t+6 u^3\Big) s^2+2 t u \Big(7 t^2+17 u t+7 u^2\Big) s+9 t^2 \
{\tpu} u^2\Big) H(1,0,1,z) \Big) \Big/ \Big( {\spt}^2 {\spu}^2 {\tpu} \Big) -4 \
H(0,y) H(1,0,1,z)-4 H(1,y) H(1,0,1,z)
\end{dmath*}
%%%% Cf^2  
\begin{dmath*}
{\white=}
+12 H(2,y) H(1,0,1,z)+8 H(3,y) \
H(1,0,1,z)-6 H(1,0,2,y)+4 H(1,z) H(1,0,3,y)+ \Big( 4 (2 s-t) t H(1,1,0,y) \Big) \Big/ \Big( s \
{\tpu} \Big) - \Big( 2 \Big(\Big(t^2-2 u t-u^2\Big) s^3+\Big(t^3-7 u t^2-2 \
u^2 t+2 u^3\Big) s^2+\Big(4 t u^3-6 t^3 u\Big) s+2 t^2 {\tpu} \
u^2\Big) H(1,1,0,z) \Big) \Big/ \Big( s {\spt}^2 {\tpu}^2 \Big) +4 H(2,y) H(1,1,0,z)-6 \
H(1,2,0,y)-4 H(1,z) H(1,2,3,y)+8 H(1,z) H(2,0,0,y)+12 H(2,0,2,y)- \Big( 2 \
\Big(\Big(6 t^2+8 u t+4 u^2\Big) s^2+u \Big(10 t^2+13 u t+7 u^2\Big) \
s+u^2 \Big(3 t^2+4 u t+3 u^2\Big)\Big) H(2,1,0,y) \Big) \Big/ \Big( {\spu}^2 \
{\tpu}^2 \Big) -4 H(1,z) H(2,1,0,y)+12 H(2,2,0,y)+16 H(1,z) H(2,2,3,y)- \Big( 2 \
\Big(8 {\tpu} s^4+\Big(17 t^2+36 u t+17 u^2\Big) s^3+\Big(9 t^3+49 \
u t^2+49 u^2 t+9 u^3\Big) s^2+2 t u \Big(10 t^2+23 u t+10 u^2\Big) s+12 \
t^2 {\tpu} u^2\Big) H(2,3,2,y) \Big) \Big/ \Big( {\spt}^2 {\spu}^2 \
{\tpu} \Big) -8 H(1,z) H(3,0,2,y)-12 H(3,0,2,y)+8 H(1,z) H(3,0,3,y)-8 H(1,z) \
H(3,2,0,y)-12 H(3,2,0,y)-24 H(3,2,2,y)+16 H(1,z) H(3,2,3,y)+8 H(1,z) \
H(3,3,0,y)+16 H(1,z) H(3,3,2,y)+24 H(3,3,2,y)-16 H(1,z) H(3,3,3,y)+8 \
H(0,0,1,1,z)+8 H(0,0,2,2,y)-8 H(0,0,3,2,y)+8 H(0,1,0,1,z)-4 H(0,1,0,2,y)+8 \
H(0,1,1,0,y)+8 H(0,1,1,0,z)-4 H(0,1,2,0,y)+8 H(0,2,0,2,y)-4 H(0,2,1,0,y)+8 \
H(0,2,2,0,y)-4 H(0,2,3,2,y)-8 H(0,3,2,2,y)+8 H(0,3,3,2,y)+4 H(1,0,0,1,z)-8 \
H(1,0,0,2,y)+4 H(1,0,1,0,y)-8 H(1,0,2,0,y)+4 H(1,0,3,2,y)+8 H(1,1,0,0,y)+8 \
H(1,1,0,1,z)+8 H(1,1,1,0,y)+8 H(1,1,1,0,z)-8 H(1,2,0,0,y)-4 H(1,2,1,0,y)-4 \
H(1,2,3,2,y)+8 H(2,0,0,2,y)-4 H(2,0,1,0,y)+8 H(2,0,2,0,y)-8 H(2,1,0,0,y)-4 \
H(2,1,0,2,y)+8 H(2,1,1,0,y)-4 H(2,1,2,0,y)+8 H(2,2,0,0,y)+16 H(2,2,3,2,y)-8 \
H(3,0,1,0,y)-8 H(3,0,2,2,y)+8 H(3,0,3,2,y)-8 H(3,2,0,2,y)+8 H(3,2,1,0,y)-8 \
H(3,2,2,0,y)+16 H(3,2,3,2,y)+8 H(3,3,0,2,y)+8 H(3,3,2,0,y)+16 \
H(3,3,2,2,y)-16 H(3,3,3,2,y)+\frac{19}{2}
\Bigg)
\Bigg\}
\end{dmath*}
%%%% Ca nf  
\begin{dmath*}
{\white=}
+
C_A n_f  \Bigg\{
\zeta_3 \Bigg(  -\frac{37}{18} \Bigg)
+
\zeta_2 \Bigg(
-\frac{1}{3} H(0,y)+\frac{1}{6} H(2,y)-\frac{1}{3} H(0,z)+\frac{1}{6} \
H(1,z)-\frac{t u}{3 s {\tpu}}-\frac{7}{36}
\Bigg)
+
\Bigg(
H(0,y) H(0,z) \Big(\frac{5}{9}-\frac{t u}{3 s {\tpu}}\Big)-\frac{t u \
H(1,0,y)}{3 s {\tpu}}+H(1,0,z) \Big(\frac{31}{36}-\frac{t u}{3 s \
{\tpu}}\Big)+\frac{(29 t+20 u) H(0,y)}{9 {\tpu}}+\frac{(20 t+29 u) \
H(0,z)}{9 {\tpu}}+\frac{31}{36} H(0,y) H(1,z)-H(0,y) \
H(0,0,z)+\frac{2}{3} H(0,y) H(0,1,z)-H(0,y) H(1,0,z)+\frac{2}{3} H(0,y) \
H(1,1,z)+\frac{31}{36} H(2,y) H(0,z)+H(2,y) H(1,z)-\frac{20}{9} H(3,y) \
H(1,z)-H(0,0,y) H(0,z)-H(0,0,y) H(1,z)-H(2,y) H(0,0,z)-\frac{2}{3} H(2,y) \
H(0,1,z)-\frac{1}{3} H(3,y) H(0,1,z)-\frac{1}{3} H(0,2,y) H(0,z)+\frac{2}{3} \
H(0,2,y) H(1,z)+H(0,3,y) H(1,z)+H(3,y) H(1,0,z)-\frac{4}{3} H(3,y) \
H(1,1,z)-H(2,0,y) H(0,z)+\frac{2}{3} H(2,0,y) H(1,z)+\frac{2}{3} H(2,2,y) \
H(0,z)-\frac{4}{3} H(2,3,y) H(1,z)+H(3,0,y) H(1,z)+H(3,2,y) \
H(0,z)-\frac{4}{3} H(3,2,y) H(1,z)-\frac{4}{3} H(3,3,y) H(1,z)+\frac{17}{27} \
H(2,y)-\frac{41}{18} H(0,0,y)+\frac{31}{36} H(0,2,y)+\frac{31}{36} \
H(2,0,y)+H(2,2,y)-\frac{20}{9} H(3,2,y)-H(0,0,2,y)-H(0,2,0,y)+\frac{2}{3} \
H(0,2,2,y)+H(0,3,2,y)-H(2,0,0,y)+\frac{2}{3} H(2,0,2,y)-\frac{2}{3} \
H(2,1,0,y)+\frac{2}{3} H(2,2,0,y)-\frac{4}{3} \
H(2,3,2,y)+H(3,0,2,y)+H(3,2,0,y)-\frac{4}{3} H(3,2,2,y)-\frac{4}{3} \
H(3,3,2,y)+\frac{17}{27} H(1,z)-\frac{41}{18} H(0,0,z)-\frac{49}{36} \
H(0,1,z)+H(1,1,z)-\frac{1}{3} H(0,0,1,z)-\frac{2}{3} \
H(0,1,1,z)-H(1,0,0,z)-\frac{2}{3} H(1,0,1,z)-\frac{439}{162}
\Bigg)
\Bigg\}
\end{dmath*}
%%%% Cf nf  
\begin{dmath*}
{\white=}
+
C_F n_f  \Bigg\{
\zeta_3 \Bigg(  -\frac{1}{9} \Bigg)
+
\zeta_2 \Bigg(
\frac{4}{3} H(1,y)-\frac{5}{3} H(2,y)-\frac{1}{3} H(1,z)+\frac{1}{3}
\Bigg)
+
\Bigg(
-\frac{(7 t+3 u) H(0,y)}{6 {\tpu}}-\frac{(3 t+7 u) H(0,z)}{6 \
{\tpu}}-\frac{31}{18} H(0,y) H(1,z)-\frac{4}{3} H(0,y) \
H(0,1,z)+\frac{1}{3} H(0,y) H(1,0,z)-\frac{4}{3} H(0,y) \
H(1,1,z)-\frac{31}{18} H(2,y) H(0,z)-2 H(2,y) H(1,z)+\frac{40}{9} H(3,y) \
H(1,z)+2 H(0,0,y) H(1,z)+2 H(2,y) H(0,0,z)+\frac{4}{3} H(2,y) \
H(0,1,z)+\frac{2}{3} H(3,y) H(0,1,z)+\frac{2}{3} H(0,2,y) H(0,z)-\frac{4}{3} \
H(0,2,y) H(1,z)-2 H(0,3,y) H(1,z)-\frac{1}{3} H(1,0,y) H(0,z)-\frac{4}{3} \
H(2,y) H(1,0,z)-2 H(3,y) H(1,0,z)+\frac{8}{3} H(3,y) H(1,1,z)+\frac{2}{3} \
H(2,0,y) H(0,z)-\frac{4}{3} H(2,0,y) H(1,z)-\frac{4}{3} H(2,2,y) \
H(0,z)+\frac{8}{3} H(2,3,y) H(1,z)-2 H(3,0,y) H(1,z)-2 H(3,2,y) \
H(0,z)+\frac{8}{3} H(3,2,y) H(1,z)+\frac{8}{3} H(3,3,y) H(1,z)-\frac{34}{27} \
H(2,y)-\frac{31}{18} H(0,2,y)+\frac{20}{9} H(1,0,y)-\frac{31}{18} H(2,0,y)-2 \
H(2,2,y)+\frac{40}{9} H(3,2,y)+2 H(0,0,2,y)-\frac{1}{3} H(0,1,0,y)+2 \
H(0,2,0,y)-\frac{4}{3} H(0,2,2,y)-2 H(0,3,2,y)-2 H(1,0,0,y)+\frac{4}{3} \
H(1,1,0,y)+2 H(2,0,0,y)-\frac{4}{3} H(2,0,2,y)-\frac{4}{3} \
H(2,2,0,y)+\frac{8}{3} H(2,3,2,y)-2 H(3,0,2,y)-2 H(3,2,0,y)+\frac{8}{3} \
H(3,2,2,y)+\frac{8}{3} H(3,3,2,y)-\frac{34}{27} H(1,z)+\frac{49}{18} \
H(0,1,z)+\frac{1}{2} H(1,0,z)-2 H(1,1,z)+\frac{2}{3} H(0,0,1,z)-\frac{1}{3} \
H(0,1,0,z)+\frac{4}{3} H(0,1,1,z)+\frac{4}{3} H(1,0,1,z)+\frac{371}{81}
\Bigg)
\Bigg\}
\end{dmath*}
%%%% Cf nf  
\begin{dmath*}
{\white=}
+
n_f^2  \Bigg\{
\frac{1}{36} H(0,y) H(0,z)-\frac{5}{27} H(0,y)+\frac{5}{36} \
H(0,0,y)-\frac{5}{27} H(0,z)+\frac{5}{36} H(0,0,z)+\frac{{\zeta_2}}{18}
\Bigg\}
\end{dmath*}
\end{dgroup*}
}

\end{appendix}

%%%%%%%%%%%%%%%%%%%%%%%%%%%%%%%%%%%%%%%%%%%%%%%%%%%%%%%%%%%%%%%%%%%%%%%%%%%%%%%%%%%%%%%%%%%

\end{document}